\documentclass[aps,prd,twocolumn,showpacs,floatfix,preprintnumbers,amsmath,amssymb,nofootinbib]{revtex4-1}

\input epsf
\usepackage{graphicx}
\usepackage{color}

\newcommand{\beq}{\begin{equation}}
\newcommand{\eeq}{\end{equation}}
\newcommand{\barr}{\begin{eqnarray}}
\newcommand{\earr}{\end{eqnarray}}

\newcommand{\kcs}{k_{\rm cs}}
\newcommand{\lcs}{\ell_{\rm cs}}

\newcommand{\physrep}{Phys. Rep.}
\newcommand{\lsim}{\mathrel{\hbox{\rlap{\lower.55ex\hbox{$\sim$}} \kern-.3em \raise.4ex \hbox{$<$}}}}
\newcommand{\gsim}{\mathrel{\hbox{\rlap{\lower.55ex\hbox{$\sim$}} \kern-.3em \raise.4ex \hbox{$>$}}}}
\begin{document}
\title{Revisiting the double-binary-pulsar probe of non-dynamical Chern-Simons gravity}

\author{Yacine Ali-Ha\"imoud} 
\email{yacine@tapir.caltech.edu}
\affiliation{California Institute of Technology, Mail Code 350-17, Pasadena, California 91125, USA}

\date{\today} 
%%%%%%%%%%%%%%%%%%%%%%%%%%%%%%%%%%%%%%%%%%%%%%%%%%%%%%%%%%%%%%%%%%%%%%%%%%%%%%%%%%%%%%%%%%%%%%%%%%
\begin{abstract}

One of the popular modifications to the theory of general relativity is non-dynamical Chern-Simons (CS) gravity, in which the metric is coupled to an externally prescribed scalar field. Setting accurate constraints to the parameters of the theory is important owing to their implications for the scalar field and/or the underlying fundamental theory. The current best constraints rely on measurements of the periastron precession rate in the double-binary-pulsar system and place a very tight bound on the characteristic CS lengthscale $k_{\rm cs}^{-1} \lesssim 3\times 10^{-9}$ km. This paper considers several effects that were not accounted for when deriving this bound and lead to a substantial suppression of the predicted rate of periastron precession. It is shown, in particular, that the point mass approximation for extended test bodies does not apply in this case. The constraint to the characteristic CS lengthscale is revised to $k_{\rm cs}^{-1} \lesssim 0.4$ km, eight orders of magnitude weaker than what was previously found.

\end{abstract}


\maketitle
\section{Introduction} 

Einstein's theory of general relativity (GR) has so far passed all observational tests with flying colors (for a review, see for example Ref.~\cite{Will_06}). It is expected, however, that GR is the low-energy limit of a more fundamental theory unifying all forces of nature. In that case it is likely that higher-order curvature corrections exist in the theory, the effect of which may become apparent in the strong field regime. Chern-Simons (CS) gravity \cite{Jackiw_Pi_03, Alexander_Yunes_09} is an example of such a higher-order modification to GR, in which the metric is coupled to a scalar field $\vartheta$ through the parity-violating Pontryagin density. In dynamical CS gravity, the scalar field is itself coupled to the metric through a wave equation sourced by the Pontryagin density. In non-dynamical CS gravity, the subject of the present work, the scalar field is externally prescribed, and is typically assumed to be a homogeneous field that only depends on cosmic time. The implications of CS gravity have been investigated in several astrophysical and cosmological scenarios. Ref.~\cite{Lue_Wang_99} considered its effects on the polarization of the cosmic microwave background, identifying the scalar field with the inflaton. Ref.~\cite{Jackiw_Pi_03} worked out the linearized theory and the propagation of gravitational waves. Ref.~\cite{Alexander_Peskin_06} suggested that CS gravity may provide a mechanism for creating the observed cosmic matter-antimatter asymmetry. Ref.~\cite{Alexander_Yunes_07D} investigated the post-Newtonian expansion of CS gravity in the point-particle limit. An important feature of the theory is that it leads to a change of frame dragging effects around rotating objects, which can be used to constrain CS gravity \cite{Alexander_Yunes_07L, Alexander_Yunes_07D}. Smith \emph{et al.} \cite{Smith_08}, hereafter SE08, have calculated the CS modification to the gravitomagnetic field around a non-relativistic, constant-density spinning body. Using the measurement of the Lense-Thirring drag around the Earth by the LAGEOS satellites \cite{Lageos}, they have set the first constraint on the characteristic CS lengthscale (defined below), $\kcs^{-1} \lesssim 1000$ km. More recently, Yunes and Spergel \cite{Yunes_Spergel_09}, hereafter YS09, used measurements of the rate of periastron precession in the double pulsar system PSR J0737-3039 A/B \cite{Burgay_03, Kramer_Wex_09} to place a stringent constraint on the CS lengthscale, $\kcs^{-1} \lesssim 3 ~\mu$m, eleven orders of magnitude stronger than the previous bound.

One of the outcomes of the calculation of SE08 is that the CS modification to the gravitomagnetic field is oscillating in space with a wavelength $2 \pi\kcs^{-1}$. This oscillatory character is due to the higher-order nature of the theory. In this paper, we show that the oscillations of the gravitomagnetic field lead to a large suppression of the periastron precession rate due to three effects that were not considered in previous studies: $(i)$ The constant-density approximation is relatively accurate for the Earth but not for a neutron star, for which the surface density is 7 orders of magnitude lower than the mean density. We show that this leads to a suppression of the gravitomagnetic field outside the source by a factor $\sim 5/(\kcs R_A)$, where $R_A$ is the radius of the rapidly rotating star A. The reason of this suppression is that less sharp boundaries result in a smaller excitation of high-frequency spatial oscillation modes. $(ii)$ The test body used for the constraint (star B) is extended, rather than a point particle. If the radius of the star $R_B$ is larger than the CS wavelength, this leads to a suppression of the average force per unit mass, by a factor $\sim 15/(\kcs R_B)^3$. $(iii)$ Even though the eccentricity of the system is $e \sim 0.09 \ll 1$ \cite{Kramer_06}, the semi-major axis $a$ is large enough that the orbital separation varies over the scale of many CS wavelengths. We show that this leads to a suppression of the secular rate of periastron precession by a factor $\sim 1/\sqrt{\kcs a e}$. Mostly because of effect $(ii)$, we expect that one cannot constrain the CS lengthscale to be much smaller than the size of a neutron star with the system considered. Properly accounting for these three effects, we indeed derive the revised constraint $\kcs^{-1} \lesssim 0.4$ km, eight orders of magnitude weaker than what was previously found, opening up the space of allowed values for the parameters of non-dynamical CS gravity. Maybe more importantly, the invalidity of the point-particle approximation for the test body shows that the ``effacing principle'' which holds in standard GR (see for example Ref.~\cite{Effacing}) is not necessarily valid in alternate theories of gravity.

This paper is organized as follows: In Sec.~\ref{sec:basics}, we review the basic equations of modified CS gravity and define our notation. We then describe the resulting gravitomagnetic field around a spinning neutron star in Sec.~\ref{sec:gravito} (the detailed calculations can be found in Appendix \ref{app:new-sol}). Sec.~\ref{sec:precession} outlines how the anomalous periastron precession rate is obtained and the effect of previous assumptions (we also provide a derivation of the rate of change of orbital elements in Appendix \ref{app:orbital}). Secs.~\ref{sec:extended} and \ref{sec:eccentricity} constitute the core of this paper, where we compute the suppression of the anomalous precession rate when correctly accounting for the finite extent of the test body and the eccentricity of the orbit, respectively. We derive the new constraint in Sec.~\ref{sec:new constraints}, and discuss how to accurately do so when the magnitude of the predicted effect is oscillatory in Appendix \ref{app:constraints}. We conclude and mention future research directions in Sec.~\ref{sec:conclusions}.

\section{Chern-Simons gravity: basic equations}\label{sec:basics} 
For a review on Chern-Simons modified gravity, we refer the reader to Ref.~\cite{Alexander_Yunes_09}. Here we simply recall the main equations and define our notation. We use geometric units throughout the paper. 

We consider the following action defining the modified theory:
\barr
S &=& \frac{1}{16 \pi}\int  R \sqrt{-g}~d^4 x + \int \mathcal{L}_{\rm mat} \sqrt{-g}~d^4 x\nonumber\\
&+& \frac{\lcs^2}{4} \int \vartheta \boldsymbol{R \tilde{R}} \sqrt{-g}~d^4 x \nonumber\\
&-& \beta \int \left[\frac12 \nabla_{\mu}\vartheta \nabla^{\mu} \vartheta+ V(\vartheta) \right] \sqrt{-g}~d^4 x.\label{eq:action}
\earr
In Eq.~(\ref{eq:action}), the first term is the standard Einstein-Hilbert action and the second term corresponds to the matter contribution with lagrangian density $\mathcal{L}_{\rm mat}$. The third term is the CS modification \cite{Jackiw_Pi_03}, which depends on the constant $\lcs^2$ (with dimensions of length squared). It couples the dimensionless CS scalar field $\vartheta$ to the metric through the Pontryagin density $\boldsymbol{R\tilde{R}}$, which is the contraction of the Riemann tensor $R_{\alpha \beta \mu \nu}$ and its dual:
\beq
\boldsymbol{R\tilde{R}} \equiv \frac12 \epsilon^{\mu \nu \sigma \tau}R^{\alpha \beta}_{~~\sigma \tau}R_{\beta \alpha \mu \nu},
\eeq
where $\epsilon^{\mu \nu \sigma \tau}$ is the four-dimensional Levi-Civita tensor. Finally, the last term of Eq.~(\ref{eq:action}) contains the canonical scalar field lagrangian, with potential $V(\vartheta)$. In the non-dynamical version of the theory initially suggested in Ref.~\cite{Jackiw_Pi_03}, $\beta = 0$. In dynamical CS theory, $\beta = 1$ (a non-zero $\beta$ can always be set to unity provided one rescales $\vartheta$, $V$ and $\lcs^2$ appropriately). These two cases constitute two qualitatively different theories.

Requesting the action to be stationary under variations of the metric results in the modified Einstein field equation:
\beq
G_{\mu \nu} + 16 \pi \lcs^2 C_{\mu \nu} = 8 \pi  \left[T_{\mu \nu}^{\rm mat} + \beta T_{\mu \nu}^{\rm \vartheta}\right], \label{eq:EF}
\eeq
where $G_{\mu \nu}$ is the Einstein tensor, $T_{\mu \nu}^{\rm mat}$ is the matter stress-energy tensor,
\beq
C^{\mu \nu} \equiv \partial_{\sigma} \vartheta ~\epsilon^{\sigma \alpha \beta (\mu } \nabla_{\alpha} R^{\nu)}_{~\beta} + \frac12 \nabla_{\tau}(\partial_{\sigma} \vartheta) ~\epsilon^{\alpha \beta \sigma (\mu} R^{\nu) \tau}_{~~~\beta \alpha}
\eeq
is a four-dimensional generalization of the Cotton-York tensor and 
\beq
T_{\mu \nu}^{\vartheta} \equiv \nabla_{\mu} \vartheta \nabla_{\nu}\vartheta - \frac12 g_{\mu\nu} \nabla_{\alpha}\vartheta \nabla^{\alpha}\vartheta - g_{\mu \nu} V(\vartheta)
\eeq
is the stress-energy tensor for the scalar field $\vartheta$.

If the scalar field is considered as dynamical, then varying the action with respect to $\vartheta$ results in the equation of motion for the scalar field:
\beq
\beta \square \vartheta =  - \frac{\ell^2_{\rm cs}}{4} \boldsymbol{R \tilde{R}} + \beta V'(\vartheta), \label{eq:theta-evol}
\eeq
where $\square$ is the usual covariant d'Alembertian operator, $\square \equiv g^{\mu \nu} \nabla_{\mu}\nabla_{\nu}$. Equation (\ref{eq:theta-evol}) can also be obtained by taking the divergence of Eq.~(\ref{eq:EF}), using the fact that $\nabla^{\mu} C_{\mu \nu} = - \frac18 \partial_{\nu}\vartheta \boldsymbol{R \tilde{R}}$, as well as the Bianchi identity $\nabla^{\mu} G_{\mu \nu} = 0$ and the conservation of the matter stress-energy tensor $\nabla^{\mu}T_{\mu \nu}^{\rm mat} = 0$.

We see that in the dynamical theory ($\beta = 1$), Eq.~(\ref{eq:theta-evol}) is an evolution equation for the scalar field, which is sourced by the Pontryagin density. In the non-dynamical theory ($\beta = 0$), Eq.~(\ref{eq:theta-evol}) is a constraint equation, $\boldsymbol{R \tilde{R}} = 0$ (the Pontryagin constraint), and the scalar field $\vartheta$ is unconstrained and needs to be externally prescribed. The Pontryagin constraint imposes a severe restriction on the set of allowed spacetimes in non-dynamical CS gravity.

In this work, we follow the approach of Refs.~\cite{Smith_08, Yunes_Spergel_09}, in that we \emph{technically} work in the dynamical theory, with simplifying assumptions that allow us to make a specific \emph{choice} for the scalar field, casting this work rather in the frame of non-dynamical CS gravity. We work in the weak field and slow rotation approximation. Outside a spinning source of angular momentum $\vec{J}$, the Pontryagin density is \cite{Yunes_Pretorius_09, Konno_09} $\boldsymbol{R \tilde{R}} = 288 M \vec{J} \cdot \vec{r} / r^8$. If we work to first order in $M/r$ and $J/r^2$, the Pontryagin density is therefore formally a second order term. For the sake of simplicity, and following SE08 and YS09, we assume a vanishing potential, $V(\vartheta) = 0$. The evolution equation for the scalar field is therefore 
\beq
\square \vartheta \approx 0,
\eeq
to lowest order in $M/r$ and $J/r^2$, i.e. $\vartheta$ is approximately a harmonic function of the coordinates. Following Refs.~\cite{Jackiw_Pi_03, Smith_08, Yunes_Spergel_09}, we make  the \emph{choice} $\vartheta = \vartheta(t) = \dot{\vartheta} t$, the assumption being that $\vartheta$ is some cosmological field that traces the evolution of cosmic time. We finally note that the energy density of the scalar field has to be much smaller than the cosmological mean density \cite{Smith_08}, and must therefore be much smaller than that of the source. We therefore have $T_{\mu \nu}^{\vartheta} \ll T_{\mu \nu}^{\rm mat}$ in Eq.~(\ref{eq:EF}). Even though we are formally working in the dynamical theory, the approximations made and our choice of scalar field, identical to the ``canonical choice'' suggested in Ref.~\cite{Jackiw_Pi_03}, therefore rather cast this work in the frame of non-dynamical CS theory, while allowing a small violation of the Pontryagin constraint (see also the discussion in Ref.~\cite{Alexander_Yunes_09}). 

With these approximations the observables only depend on the combination $ \lcs^2 \dot{\vartheta}$, which has dimensions of length. We define the characteristic CS wavenumber, which has units of inverse length:
\beq
\kcs \equiv (8 \pi \lcs^2 \dot{\vartheta})^{-1}.
\eeq
The correspondence with the notation of Refs.~\cite{Smith_08, Yunes_Spergel_09} is $\kcs = m_{\rm cs} = 2/\tau_{\rm CS}$.

We emphasize that we have made several simplifying and somewhat arbitrary choices (following previous works), so that one should consider the theory studied as a toy model, aimed at gaining some insight into the more complex underlying fundamental physics.

\section{Gravitomagnetic field around a spinning object} \label{sec:gravito}

We follow SE08 and work in the non-relativistic, slow rotation regime. We consider the stationary (time-independent) problem. We work with the usual gravitomagnetic vector potential $\vec{A}$, in the Coulomb gauge $\vec{\nabla}\cdot \vec{A} = 0$ [gauge freedom on spatial coordinates allows this choice; in the full time-dependent problem gauge freedom is limited by the choice of $\vartheta$ which determines the time coordinate]. With this setting, SE08 have shown that the only difference between CS gravity and GR is Ampere's law, which takes the form
\beq
\nabla^2\left[\vec{A} + \kcs^{-1} \vec{\nabla} \times \vec{A}\right] = - 4 \pi \vec{j},\label{eq:Ampere2}
\eeq
or, equivalently, in terms of the gravitomagnetic field $\vec{B} \equiv \vec{\nabla} \times \vec{A}$,
\beq
\vec{\nabla} \times \vec{B} - \kcs^{-1} \nabla^2 \vec{B} = 4 \pi \vec{j}, \label{eq:Ampere}
\eeq
where $\vec{j}$ is the mass current of the source. The standard GR equation is recovered for $\kcs \rightarrow \infty$. For non-zero $\kcs^{-1}$, Eq.~(\ref{eq:Ampere2}) is a third order partial differential equation, and its solutions are therefore qualitatively different than those of GR.

SE08 have solved Eq.~(\ref{eq:Ampere2}) and computed the resulting gravitomagnetic field inside and outside a constant density object. This is appropriate for the Earth, in which the density varies by a factor of a few from the center to the edge. This is however not accurately representing a neutron star, where the density varies smoothly from the core to the edge, where it is $\sim 10^{-7}$ smaller than the mean density. This smooth edge decreases the CS-induced gravitomagnetic field outside the star for large $\kcs R$, as the high-spatial-frequency modes pick up a smaller amplitude. In Appendix \ref{app:new-sol}, we solve the modified Ampere's equation for a more realistic neutron star density profile:
\beq
\rho(r) = \rho_c\left[ 1 - (r/R)^2\right]. \label{eq:rho-realistic}
\eeq
We compute the gravitomagnetic field and obtain that the CS modification $\vec{B}_{\rm CS} \equiv \vec{B} - \vec{B}_{\rm GR}$, where $\vec{B}_{\rm GR}$ is the gravitomagnetic field for standard GR, is of the form \cite{Smith_08}
\barr
\vec{B}_{\rm CS} &=& 4 \pi \overline{\rho} R^2 \Big{\{} D_1(r) \vec{\Omega} + D_2(r) \hat{r} \times \vec{\Omega} \nonumber\\
&& ~~~ ~~~~~~~~~ + D_3(r) \hat{r} \times (\hat{r} \times \vec{\Omega})\Big{\}}, \label{eq:BCS}
\earr
where $\overline{\rho} = 3M/(4 \pi R^3)$ is the mean density of the star. Outside the star, the functions $D_i$ are rescaled from the result of SE08 by a factor of $5j_3(\kcs R)/[\kcs R j_2(\kcs R)]$ and are given by
\barr
D_1(r \geq R) &=& 10 j_3(\kcs R)  \frac{y_1(\kcs r)}{\kcs r}, \label{eq:D1-main}\\
D_2(r \geq R) &=& 5 j_3(\kcs R) y_1(\kcs r), \\
D_3(r \geq R) &=& 5 j_3(\kcs R) y_2(\kcs r), \label{eq:D3-main}
\earr
where $j_n$ and $y_n$ are the order-$n$ spherical Bessel functions of the first and second kind, respectively. 
We find that for $\kcs R \gg 1$, the envelope of the oscillating gravitomagnetic field outside the star is reduced by a factor $5/(\kcs R)$ when using the density profile (\ref{eq:rho-realistic}) rather than assuming a constant density. We will show in Sec.~\ref{sec:new constraints} that our final constraint is weakly dependent on the exact density profile and the simple form (\ref{eq:rho-realistic}) is sufficiently accurate for the purpose of this calculation.

\section{Anomalous periastron precession in CS gravity}\label{sec:precession}

In the binary pulsar, the orbital angular momentum is nearly aligned with the spin vector $\vec{J}$ of the central rotating source \cite{Ferdman_08}, as one may expect from formation mechanisms. With this geometry, the CS gravitomagnetic field leads to an anomalous secular precession of the periastron with a rate (see Appendix \ref{app:orbital} for a derivation):
\beq
\dot{\omega}_{\rm CS} = \frac{4}{a^3(1 - e^2)^{3/2}}  \big{\langle} r^3\overline{B}_0(r) [2 + (e^{-1} + e) \cos f] \big{\rangle}_f, \label{eq:omega_dot}
\eeq
where $B_0(r) \equiv 4 \pi \overline{\rho} R^2 [D_1(r) - D_3(r)]$, $f$ is the true anomaly of the Keplerian trajectory, $r(f)$ is the distance from the central rotating object and has the standard form given by Eq.~(\ref{eq:r(f)}) for an ellipse of semi-major axis $a$ and eccentricity $e$, and $\langle X \rangle_f$ is the angle-weighted average of $X$ over one orbit [see Eq.~(\ref{eq:<X>f})]. Moreover, we have used $\overline{B}_0(r)$, the mass-weighted average of the gravitomagnetic field inside the test body (we will get back to this point shortly).

In SE08 and YS09, the two following assumptions were implicitly made:

$(i)$ The test body was assumed to be a point mass, so that one can use $\overline{B}_0(r) = B_0(r)$ in Eq.~(\ref{eq:omega_dot}).

$(ii)$ The eccentricity of the orbit was assumed to be small enough that the radius could be expanded around the semi-major axis inside the angular averaging, $r \approx a(1 - e \cos f)$ and in particular $B_0(r) \approx B_0(a) - a e B_0'(a)\cos f$. 

With these assumptions, one would obtain, to lowest order in eccentricity, $\dot{\omega}_{\rm CS} = 2[B_0(a) - a B_0'(a)]$. 

\begin{figure}
\includegraphics[width = 85 mm]{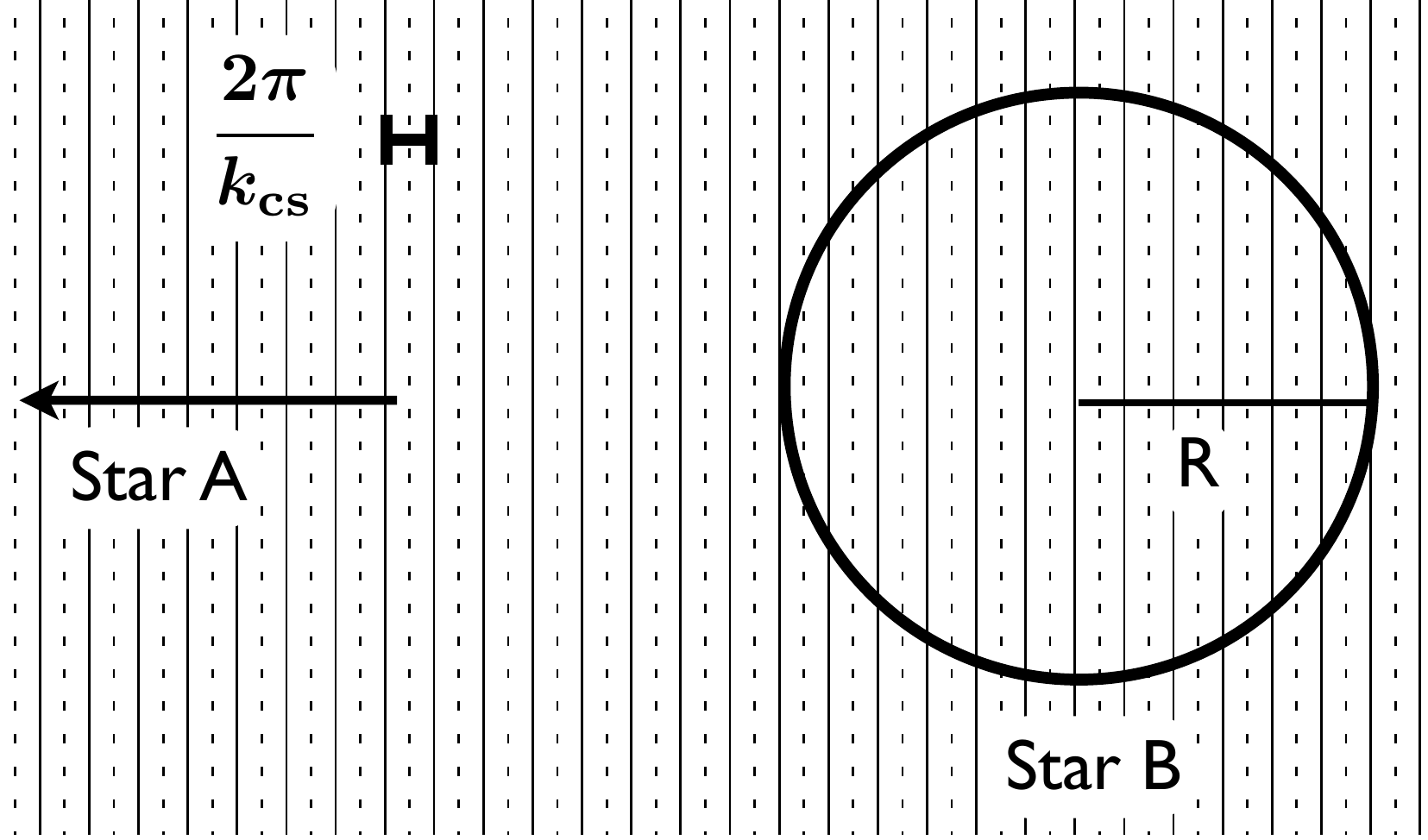}
\caption{Schematic representation (not to scale) of the first effect discussed in this paper. The solid and dotted lines represent the crests and troughs of the oscillating gravitomagnetic field, with wavelength $2 \pi \kcs^{-1}$. The test body, star B, extends over many wavelengths of the CS gravitomagnetic field. The force per unit mass is therefore largely suppressed compared to the case of a point particle.} \label{fig:extended} 
\end{figure}

Assumption $(i)$ is only valid if the size of the test body is much smaller than the wavelength of the oscillating CS gravitomagnetic field, $2 \pi \kcs^{-1}$. This is indeed the case for the analysis done by SE08, who used the LAGEOS satellites \cite{Lageos} as test bodies, and who derived a relatively weak constraint $\kcs^{-1} \lesssim 1000$ km. In this work as in YS09, the test body is star B of the double-binary pulsar, with radius $R_B \sim 10$ km. YS09 derive a constraint $\kcs^{-1} \lesssim 3 \ \mu$m, many orders of magnitude smaller than the size of the test body. In that case, the CS gravitomagnetic force $\vec{f}_{\rm CS} = - 4 \vec{v} \times \vec{B}_{\rm CS}$ oscillates a large number of times within the test body, resulting in a near cancellation of the average force per unit mass. We illustrate this effect schematically in Fig.~\ref{fig:extended}. We will show in Sec.~\ref{sec:extended} that the mass-averaged gravitomagnetic field in star B is
\beq
\overline{B}_0(r) = 15 \frac{j_2(\kcs R_B)}{(\kcs R_B)^2} B_0(r), \label{eq:B0-supp}
\eeq
which is suppressed by a factor $\sim 15/(\kcs R_B)^3$ from the point-mass case when $\kcs R_B \gg 1$.

\begin{figure}
\includegraphics[width = 85 mm]{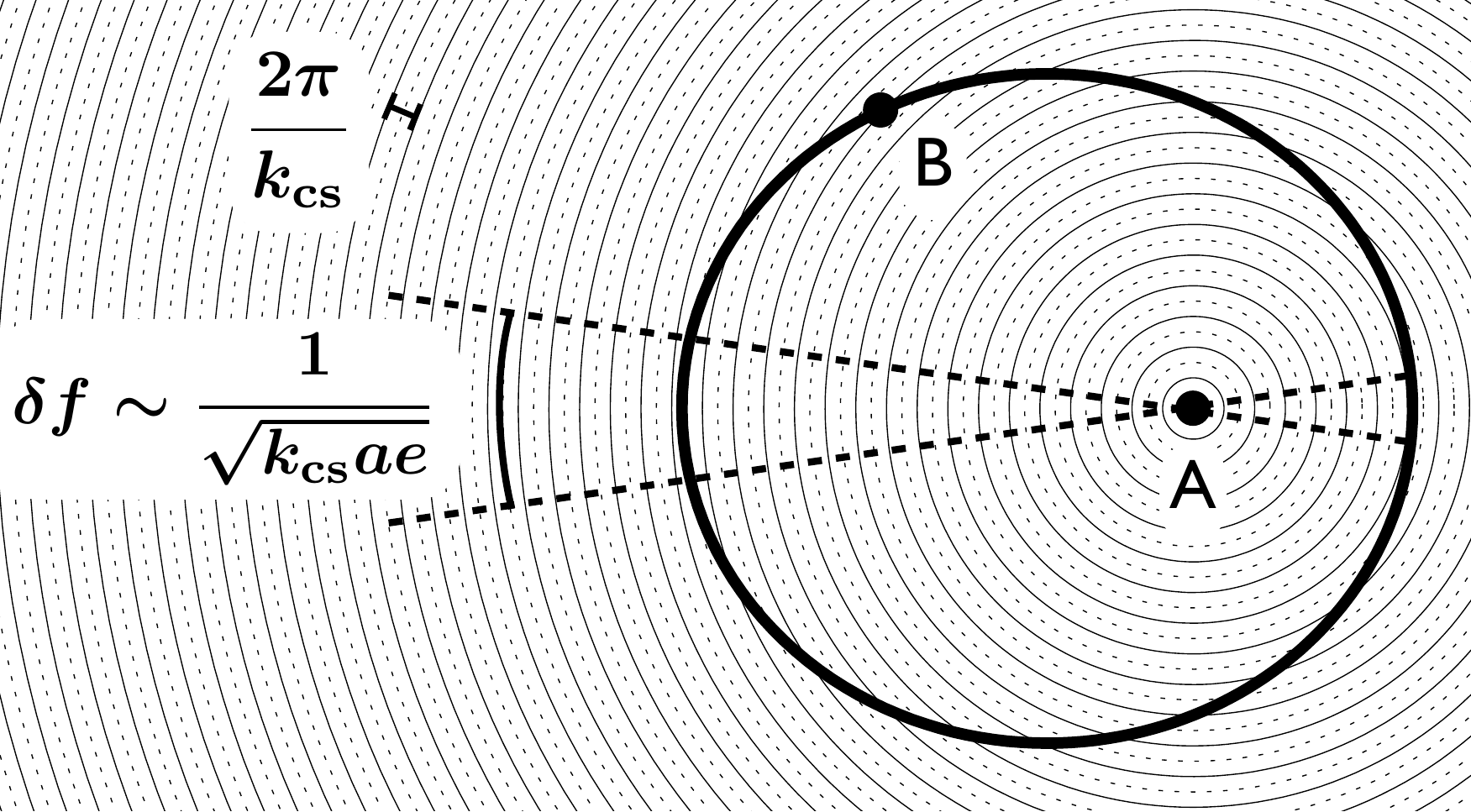}
\caption{Schematic representation (not to scale) of the second effect discussed in this paper. The solid and dotted lines represent the crests and troughs of the oscillating gravitomagnetic field, with wavelength $2 \pi \kcs^{-1}$. The figure represents the orbit as seen face-on. During its eccentric trajectory, star B goes through many wavelengths of the oscillating CS gravitomagnetic field generated by the rotating star A. The net force nearly averages out, except for a small section at the pericenter and apocenter. Note that the scale used here is different from that of Fig.~\ref{fig:extended}.} \label{fig:eccentric} 
\end{figure}

Let us now examine assumption $(ii)$. The orbital separation of the binary varies within the range $a  - a e \leq r \leq a + ae$, i.e. the difference between the radial separation at apoastron and periastron is $2 ae$. This corresponds to $2a e \kcs/(2 \pi)$ oscillations of the CS gravitomagnetic field. In the case studied by SE08, $a \sim 12000$ km, $e < 0.01$ and therefore $2 a e\ll 2 \pi \kcs^{-1}$ when the constraint of SE08 is saturated, i.e. the test body remains on roughly the same wavefront of the gravitomagnetic field throughout its orbit. In the double-binary-pulsar system, however, $a \approx 4\times 10^5$ km and $e \approx 0.09$ \cite{Kramer_Wex_09}, and therefore $2 a e \approx 7 \times 10^4$ km $ \gg \kcs^{-1}$, since SE08 have shown that $\kcs^{-1} \lesssim $ 1000 km. The test body therefore goes through many oscillations of the gravitomagnetic field during each orbit, and the resulting force nearly averages out. We illustrate this effect in Fig.~\ref{fig:eccentric}. Mathematically speaking, since $\overline{B}_0(r) \propto \cos(\kcs r)$, one \emph{cannot} approximate $\overline{B}_0(r) \approx \overline{B}_0(a) - a e B_0'(a) \cos f$ \emph{unless} $\kcs a e \ll 1$. We will show in Sec.~\ref{sec:eccentricity} that properly averaging the gravitomagnetic force on an orbit leads to an additional suppression of the secular rate of periastron precession by a factor $\sim 1/(e\sqrt{\pi \kcs a e})$.

Before turning to the computation of the effect mentioned above, we emphasize that these two effects are \emph{additive}, i.e. the mass-averaged gravitomagnetic field $\overline{B}_0(r)$ remains a rapidly oscillating function of the position of the test body's center of mass [it is just uniformly suppressed by a constant factor as can be seen in Eq.~(\ref{eq:B0-supp})]. 


\section{Average force per unit mass for an extended body} \label{sec:extended}

In standard GR, all metric components are slowly varying functions of position in the far-field of a given source. As a consequence, the motion of an extended test body of characteristic size $R$ at a distance $r \gg R$ from a central source can be approximated by the motion of its center of mass, regardless of its internal structure. The latter only appears in corrections of relative amplitude $(R/r)^2$. This ``effacement'' of internal structure \cite{Effacing}, however, does not necessarily apply in alternative theories of gravitation \cite{Will_1981}, in particular in non-dynamical CS gravity, as we shall show in this section.

Consider a spherically symmetric test body of radius $R$, mass $M$ and mass density $\rho$. The spherical Bessel functions can be written in terms of simple trigonometric functions, and the local CS gravitomagnetic force $\vec{f}_{\rm CS} = -4 \vec{v} \times \vec{B}_{\rm CS}$ is of the form 
\beq
f_{\rm CS}(r) = f_c(r) \cos(\kcs r) + f_s(r) \sin (\kcs r),
\eeq
where $f_c(r)$ and $f_s(r)$ are slowly varying functions of $r$ (they are or the form $1/r^n$). Denoting $\vec{r}_0$ the position of the test body's center, the average force per unit mass on the entire body is
\barr
\overline{f}_{\rm CS}(r_0) &\equiv& \frac1M \int_{\mathcal{V}} \rho(r') f\left( |\vec {r}_0 + \vec{r}~'|\right) d^3r' \nonumber\\
&\approx& \frac1M f_c(r_0) \int_{\mathcal{V}} \rho(r') \cos\left(\kcs |\vec {r}_0 + \vec{r}~'|\right) d^3r' \nonumber\\
&+& \frac1M f_s(r_0) \int_{\mathcal{V}} \rho(r') \sin\left(\kcs |\vec {r}_0 + \vec{r}~'|\right) d^3r'. 
\earr 
where $\mathcal{V}$ is the body's volume, and since the amplitudes $f_c$ and $f_s$ remain nearly constant over the body's extent (if $R\ll r_0$), we can take them out of the integrals. We define the projected linear density
\beq
\lambda(r') \equiv \int_0^{\sqrt{R^2-r'^2}} \rho\left(\sqrt{r'^2 + y^2}\right) 2 \pi y d y.
\eeq
Neglecting the curvature of planes of constant phase (valid if $\kcs a \ll a^2/R^2$, indeed satisfied when our constraint is saturated), we rewrite the average force per unit mass as 
\barr
\overline{f}_{\rm CS}(r_0) &=&  \frac{1}{M} f_c(r_0)\int_{-R}^R \lambda(r') \cos[\kcs(r_0 + r')] dr'\nonumber\\
&+& \frac{1}{M} f_s(r_0)\int_{-R}^R \lambda(r') \sin[\kcs(r_0 + r')] dr' 
\earr
Expanding the trigonometric functions and using the fact that $\lambda$ is an even function, we arrive at
\beq
\overline{f}_{\rm CS}(r_0) = \mathcal{I}(\kcs R) f_{\rm CS}(r_0),
\eeq
where we have defined
\beq
\mathcal{I}(\kcs R) \equiv \frac2M \int_{0}^R \lambda(r') \cos(\kcs r') d r'. \label{eq:I-def}
\eeq
If the body's extent is much less than a CS wavelength, $\kcs R \ll 1$, and $\mathcal{I}(\kcs R) = 1$ so
\beq
\overline{f}_{\rm CS}(r_0; \kcs R \ll 1) = f_{\rm CS}(r_0).
\eeq
This assumption was implicitly made in Refs.~\cite{Smith_08, Yunes_Spergel_09} to compute the force acting on the test body.

For a general value of $\kcs R$, however, $\mathcal{I}(\kcs R) \neq 1$, and in particular, for $\kcs R \gg 1$, the rapidly oscillating integrand in Eq.~(\ref{eq:I-def}) leads to a large suppression of $\mathcal{I}$. We start by rewriting $\mathcal{I}$, after integrating by parts and noticing that $\lambda(R) = 0$ and $d\lambda/d r' = - 2 \pi r' \rho(r')$:
\barr
\mathcal{I}(\kcs R) &=& \frac{1}{\kcs R}\frac{4 \pi R}{M}\int_{0}^R  r' \rho(r') \sin(\kcs r') d r'\nonumber\\
                          &=& \frac3{\kcs R} \int_0^1 x \tilde{\rho}(x) \sin(\kcs R x) dx,    \label{eq:Integral}
\earr
where we have defined the normalized density profile $\tilde{\rho}(r') \equiv \rho(r')/\overline{\rho}$ and made the change of variable $x = r'/R$. For an approximate neutron star density profile as in Eq.~(\ref{eq:rho-realistic}), we obtain
\beq
\mathcal{I}(\kcs R) = 15 \frac{j_2(\kcs R)}{(\kcs R)^2}.
\eeq
We see that for $\kcs R \gg 1$, $\mathcal{I} \sim 15/(\kcs R)^3$ and the average force per unit mass is largely suppressed. Note that even for a constant density profile (and therefore sharp edges), one would obtain $\mathcal{I}(\kcs R) = 3 j_1(\kcs R)/(\kcs R) \sim 3/(\kcs R)^2$ for $\kcs R \gg 1$, i.e. the main cause of the suppression is the fact that the test body is extended, and the smooth boundary leads to an additional suppression by a factor $\sim 5/(\kcs R)$, just as in the case of the rotating neutron star that sources the CS gravitomagnetic field. 

We therefore conclude that the average force per unit mass on neutron star $B$, with radius $R_B$ and situated at a distance $r$ from the rotating body (star A) is
\beq
\overline{f}_{\rm CS}(r) = 15 \frac{j_2(\kcs R_B)}{(\kcs R_B)^2} f_{\rm CS}(r).
\eeq
This result can equivalently be quoted in terms of the mass-averaged gravitomagnetic field, Eq.~(\ref{eq:B0-supp}), since $\vec{f}_{\rm CS} = - 4 \vec{v} \times \vec{B}_{\rm CS}$ and $\vec{v}$ is a slowly varying function of position.

\section{Effect of eccentricity on the secular change of orbital elements}\label{sec:eccentricity}

Since we already know from Solar System constraints \cite{Smith_08} that $\kcs a \gtrsim 400 \gg 1$ for the system considered, we can first expand the spherical bessel functions in $\kcs r \gg 1$. Using Eqs.~(\ref{eq:B0-supp}) and (\ref{eq:D1-main})-(\ref{eq:D3-main}), with 
\beq
B_0(r) = \frac{3 M_A}{R_A} \Omega_A [D_1(r) - D_3(r)],
\eeq
we arrive at
\beq
\overline{B}_0(r) \approx \mathcal{B}_0\frac{\cos(\kcs r)}{\kcs r},\label{eq:B0-final}
\eeq
where we have defined the amplitude
\beq
\mathcal{B}_0 \equiv -225 \frac{M_A}{R_A} \Omega_A j_3(\kcs R_A) \frac{j_2(\kcs R_B)}{(\kcs R_B)^2}. 
\eeq
We now have all elements at hand to compute the rate of periastron precession. Using Eq.~(\ref{eq:B0-final}) in Eq.~(\ref{eq:omega_dot}), we obtain, to lowest order in eccentricity:
\beq
\dot{\omega}_{\rm CS} = \frac{4 \mathcal{B}_0} {\kcs a e} \langle \cos[\kcs r(f)] \cos f \rangle_f. 
\eeq
If it were not for the rapidly oscillating function, this term would in fact be independent of eccentricity to lowest order in $e$; however, we will see that for $\kcs a e \gg 1$, the $1/e$ dependence remains so this is the dominant term in Eq.~(\ref{eq:omega_dot}).

In the small eccentricity regime $e \ll 1$, we have $r(f) \approx a (1 - e \cos f)$ and therefore
\barr
\langle \cos[\kcs r(f)] \cos f\rangle_f &\approx& \sin(\kcs a) \langle \sin(\kcs a e \cos f) \cos f \rangle_f\nonumber\\
 &=& \sin(\kcs a) J_1(\kcs a e), \label{eq:lim-e-0}
\earr
where $J_1$ is the order-1 Bessel function of the first kind. In Eq.~(\ref{eq:lim-e-0}) we have expanded $\cos[r(f)]$ and used the fact that the term proportional to $\cos(\kcs a e \cos f) \cos f$ averages to zero (it flips sign under the change $f \rightarrow \pi - f$). Strictly speaking, Eq.~(\ref{eq:lim-e-0}) is only valid if $\kcs a e^2 \ll 1$, required to approximate $\cos[\kcs r(f)] \approx \cos[\kcs a (1 - e \cos f)]$; the final result is however independent of the order with which we take the limits $\kcs a \gg 1$, $e \ll 1$. Now expanding the Bessel function for $\kcs a e \gg 1$, we finally obtain
\beq
\dot{\omega}_{\rm CS} \approx - \frac{4 \mathcal{B}_0} {\kcs a e} \sin(\kcs a) \frac{\sqrt{2} \cos(\kcs a e + \frac{\pi}4)}{\sqrt{\pi \kcs a e}}.
\eeq
We see that the rate of periastron precession is suppressed by a factor $\sim 1/\sqrt{\pi \kcs a e}$ for $\kcs a e \gg 1$. 

We can understand the magnitude of the suppression with the following simple argument. When $\kcs a e \gg 1$, the test body goes through a large number of peaks and troughs of $\vec{B}_{\rm CS}$ during an orbit, the effect of which averages out nearly exactly. The only locations where the test body remains on the same wavefront of the oscillatory metric is at the pericenter and apocenter of the trajectory, where the orbit osculates a circle (see Fig.~\ref{fig:eccentric}). We expand $r(f) \approx a ( 1- e \cos f)$ near the pericenter ($ r_{-} = a(1-e), \ f_- = 0$) or apocenter ($r_+ = a(1 + e), \ f_+ = \pi$) and obtain
\beq
r - r_{\pm}  \approx \mp a e \frac{(f - f_{\pm})^2}{2}.
\eeq 
The test body remains on the same wavefront of $\vec{B}_{\rm CS}$ as long as $\kcs|r - r_{\pm}| \lesssim \pi/2$, which translates to
\beq
|f - f_{\pm}| \lesssim \delta f \equiv \sqrt{\frac{\pi}{\kcs a e}}. 
\eeq  
Therefore the CS-induced forces do not average out only on a fraction of the orbit
\beq
\frac{4\delta f}{2 \pi} \sim \frac{2}{ \sqrt{\pi \kcs a e}},
\eeq
and the secular variation of orbital parameters will be reduced by this factor with respect to a perfectly circular orbit.

\section{Revised constraints} \label{sec:new constraints}
We can now use the measured periastron precession rate in the binary pulsar to set a constraint on the CS length $\kcs^{-1}$. Anticipating that the constraint will still be such that $\kcs^{-1} \ll 10$ km, we can expand $\mathcal{B}_0$ in $\kcs R_A \gg 1, \ \kcs R_B \gg 1$. Our final result for the rate of periastron precession is therefore
\barr
\dot{\omega}_{\rm CS} &\approx& -900 \sqrt{\frac2{\pi}} \frac{M_A}{R_A}\Omega_A \sin(\kcs a) \frac{\cos(\kcs a e + \frac{\pi}4)}{(\kcs a e)^{3/2}}\nonumber\\
&\times& \frac{\cos(\kcs R_A)}{\kcs R_A} \frac{\sin(\kcs R_B)}{(\kcs R_B)^3}.\label{eq:omega.dot.final}
\earr
This is an oscillating function of $\kcs$, which adds a little subtlety to the meaning of constraints (we discuss this point in Appendix \ref{app:constraints}). As a simple estimate, we can approximate $\dot{\omega}_{\rm CS}$ by its envelope (i.e. replace all the sinusoidal functions by unity):
\beq
\dot{\omega}_{\rm CS} \sim 900 \sqrt{\frac2{\pi}} \frac{M_A}{R_A}\Omega_A (\kcs a e)^{-3/2}(\kcs R_B)^{-4}.
\eeq 

If one can measure the rate of periastron precession to be within an error $\delta \dot{\omega}$ from the value predicted by GR, then we can constrain the CS length to be less than
\beq
\kcs^{-1} \lesssim \left( \frac{1}{900} \sqrt{\frac{\pi}{2}}\frac{\delta \dot{\omega}}{\Omega_A} \frac{R_A}{M_A}  R_B^4 (ae)^{3/2}\right)^{1/5.5}. \label{eq:constraint}
\eeq
With $R_B \sim 10$ km, $M_A/R_A \sim 1/5$, $a \sim 4 \times 10^5$ km, $e \sim 0.09$ and $2\pi/\Omega_A \approx 22$ ms, \cite{Kramer_Wex_09} and assuming (as in YS09) that the measured precession rate agrees with the GR prediction within the measurement error $\delta \dot{\omega} \approx 0.05$ degrees per year \cite{Kramer_06}, we obtain 
\beq
\kcs^{-1} \lesssim 0.2 ~\textrm{km} \left(\frac{\delta \dot{\omega}}{0.05 \ \textrm{deg/yr}}\right)^{1/5.5}.
\eeq
This constraint is eight orders of magnitude weaker than what was found by YS09, but remains more than three orders of magnitude stronger than what can be inferred from Solar System tests. Note that because of the small exponent in Eq.~(\ref{eq:constraint}), the constraint depends only weakly of the exact value chosen for $\delta \dot{\omega}$ as well as on the exact value of the semi-major axis and eccentricity. Accounting more precisely for the oscillatory amplitude in Eq.~(\ref{eq:omega.dot.final}) leads to the formally more accurate constraint (see Appendix \ref{app:constraints} for details):
\beq
\kcs^{-1} < 0.4 \ \textrm{km} \ \ \ [85\% \textrm{ confidence level}], \label{eq:constraint-final}
\eeq
in the sense that there is still a 15\% chance that $\kcs^{-1} \gtrsim 0.4$ km but the predicted $\dot{\omega}_{\rm CS}$ remains consistent with observational constraints due to a low amplitude of the oscillatory functions in Eq.~(\ref{eq:omega.dot.final}).

Finally, one might worry that the simple density profile that we have used for both neutron stars may not capture surface effects; however, for $\kcs \sim 0.4$ km, the size of the star is only a few times the CS wavelength and a coarse-grained density profile is largely sufficient. For example, using an even simpler constant density profile for both stars would only change the constraint by a factor of $\sim 2.5$, making it tighter. We also explicitly evaluated the integral (\ref{eq:Integral}) for realistic density profiles computed with physically motivated equations of state (EOS) \cite{Evan_thanks}, and obtain $\mathcal{I} \sim 3/(\kcs R)^{2.5}$ for a wide range of neutron star masses and independently of the EOS chosen (the EOSs used where those of Refs.\cite{LS_EOS} and \cite{Shen_1, Shen_2}). We therefore conclude that using the density profile (\ref{eq:rho-realistic}) provides a conservative, yet relatively accurate constraint, and we adopt Eq.~(\ref{eq:constraint-final}) as our final result.

\section{Conclusions} \label{sec:conclusions}

In this work, we have re-evaluated the correction to the rate of periastron precession in the double-binary-pulsar induced by the Chern-Simons modification to general relativity. We found it to be largely suppressed compared to previous results, where several important effects were not accounted for. This large suppression is due to the oscillatory character of the CS gravitomagnetic field which leads to near cancellations of the mean force inside an extended test body and during an orbit, as well as a reduction of the modified gravitomagnetic field generated by the source if its boundary is not sharp. We have revised the constraints on the CS characteristic length accordingly: $\kcs^{-1} \equiv (8 \pi \lcs^2 \dot{\vartheta})^{-1} \lesssim 0.4$ km. This new constraint opens up the space of allowed values for the coupling strength of the CS theory $\ell_{\rm cs}^2$ or the scalar field derivative $\dot{\vartheta}$. One should keep in mind, however, that we have made several simplifying choices for the parameters of the theory (see Sec.~\ref{sec:basics}), and this result only applies to the very specific simplified theory that was studied in this work.

Finally, we would like to point out that a unique feature of the CS gravitomagnetic field is the presence of a poloidal component, the effects of which have not been fully investigated yet. We show in Appendix \ref{app:poloidal} that it results in very interesting orbital dynamics. In particular, it leads to a change of the angle between the orbital angular momentum and the spin of the source, as well as secular changes in the eccentricity and the magnitude of the orbital angular momentum. Such effects are \emph{qualitatively} different than standard GR predictions and could provide a powerful handle to constrain CS gravity further. A full study of the resulting dynamics would require including simultaneously other spin-orbit coupling effects in GR and CS gravity \cite{Alexander_Yunes_07D}, and will be the subject of future work.

\acknowledgments{The author is indebted to Christopher Hirata for making many insightful comments on this work and pointing out the additional weakening of the constraint due to the non-sharp boundary of neutron stars. The author also thanks Daniel Grin and Nico Yunes for a careful reading of the draft of this paper and making valuable comments, and Yanbei Chen, Marc Kamionkowski, Tristan Smith, Adrienne Erickcek and Frans Pretorius for useful discussions on Chern-Simons gravity. This work was supported by the U.S. Department of Energy (Contract No. DE-FG03-92-ER40701) and the National Science Foundation (Contract No. AST-0807337).}

\appendix

\section{Solution to the modified Ampere's equation in CS gravity} \label{app:new-sol}

In this Appendix we derive a solution to Ampere's equation in CS gravity, Eq.~(\ref{eq:Ampere2}). Such a solution was already provided in SE08 for the case of a constant density body in solid rotation. We consider more general density profiles, still assuming solid rotation with constant angular velocity $\vec{\Omega}$. 

We start by writing the vector potential $\vec{A} = \vec{A}_{\rm GR} + \vec{A}_{\rm CS}$, where $\vec{A}_{\rm GR}$ is the solution of Ampere's law in standard GR, 
\beq
\nabla^2 \vec{A}_{\rm GR} = - 4 \pi \vec{j}. \label{eq:Ampere_GR}
\eeq
Equation (\ref{eq:Ampere2}) becomes, after multiplying by $\kcs$:
\beq
\nabla^2 \left[\vec{\nabla}\times \vec{A}_{\rm CS} + \kcs \vec{A}_{\rm CS} + \vec{\nabla} \times \vec{A}_{\rm GR}\right] = \vec{0}, 
\eeq
Requiring the term in brackets to vanish at infinity, the only solution for this Laplace's equation is 
\beq
\vec{\nabla}\times \vec{A}_{\rm CS} + \kcs \vec{A}_{\rm CS} = - \vec{\nabla}\times \vec{A}_{\rm GR}, \label{eq:Ampere_CS}
\eeq
or equivalently,
\beq
\vec{B}_{\rm CS} + \kcs \vec{A}_{\rm CS} = - \vec{B}_{\rm GR}, \label{eq:Ampere_CS2}
\eeq
where $\vec{B}_{\rm GR} = \vec{\nabla}\times \vec{A}_{\rm GR}$ and $\vec{B}_{\rm CS} = \vec{\nabla}\times \vec{A}_{\rm CS}$.
We now work in spherical polar coordinates $r, \theta, \phi$, choosing $\vec{\Omega}$ as the $z$-axis. The associated unit vectors are $\hat{e}_{r} \equiv \hat{r}$, $\hat{e}_{\theta} \equiv (\sin \theta)^{-1} \hat{r} \times (\hat{r} \times \hat{\Omega})$ and $\hat{e}_{\phi} \equiv (\sin \theta)^{-1} \hat{\Omega} \times \hat{r}$. We decompose a given vector $\vec{V}$ on this basis as $\vec{V} = V^{\hat r} \hat{e}_r + V^{\hat \theta} \hat{e}_{\theta} + V^{\hat \phi}\hat{e}_{\phi}$. Since the GR gravitomagnetic field does not have any poloidal component ($B_{\rm GR}^{\hat \phi} = 0$), projecting Eq.~(\ref{eq:Ampere_CS2}) on $\hat{e}_{\phi}$ gives us a first relation
\beq
B_{\rm CS}^{\hat \phi} = - \kcs A_{\rm CS}^{\hat \phi}. \label{eq:BCSphi}
\eeq
Writing explicitly the curl in spherical polar coordinates, we also obtain (using azimuthal symmetry)
\barr
B_{\rm CS}^{\hat r} &=& \frac{1}{r \sin \theta}\frac{\partial}{\partial \theta}\left(\sin \theta  A_{\rm CS}^{\hat \phi}\right),\\
B_{\rm CS}^{\hat \theta} &=& -\frac{1}{r} \frac{\partial}{\partial r}\left(r A_{\rm CS}^{\hat \phi}\right). \label{eq:BCStheta}
\earr
We therefore see that the CS correction to the gravitomagnetic field is entirely determined by the poloidal component of the CS vector potential, $A_{\rm CS}^{\hat \phi}$. We now multiply Eq.~(\ref{eq:Ampere_CS}) by the operator $(\kcs - \vec{\nabla} \times)$. The Coulomb gauge implies $\vec{\nabla} \times \vec{\nabla} \times \vec{A} = - \nabla^2 \vec{A}$, and using Eq.~(\ref{eq:Ampere_GR}), we see that $\vec{A}_{\rm CS}$ is the solution of the inhomogeneous Helmholtz equation (similar to that obtained in SE08):
\beq
\nabla^2 \vec{A}_{\rm CS} + \kcs^2 \vec{A}_{\rm CS} = 4 \pi \vec{j} - \kcs \vec{B}_{\rm GR}. \label{eq:Helmholtz}
\eeq
We now project Eq.~(\ref{eq:Helmholtz}) on $\hat{e}_{\phi}$ and obtain the following equation for $A_{\rm CS}^{\hat \phi}$:
\barr
&&\Bigg{\{}\frac{\partial^2}{\partial r^2} + \frac{2}r \frac{\partial}{\partial r} + \kcs^2  \nonumber\\
&&+ \frac{1}{r^2 \sin \theta} \left(\sin \theta \frac{\partial^2}{\partial \theta^2} + \cos \theta \frac{\partial}{\partial \theta} - 1\right)\Bigg{\}} A_{\rm CS}^{\hat \phi} \nonumber\\
&&= 4 \pi r \rho(r) \Omega \sin \theta,
\earr
where we used $\vec{j} = \rho(r) \vec{\Omega} \times \vec{r}$. This equation is separable, and the solution is of the form 
\beq
A_{\rm CS}^{\hat \phi}(r, \theta) = 4 \pi \overline{\rho} R^3 \Omega \sin \theta \mathcal{A}\left(\frac{r}{R}; \kappa\right), \label{eq:separable}
\eeq 
where $\overline{\rho} = 3 M/(4 \pi R^3)$ is the mean density of the star. The function $\mathcal{A}(x)$ [with $x = r/R$] in Eq.~(\ref{eq:separable}) is the solution of the inhomogeneous spherical Bessel equation
\beq
\mathcal{A}''(x) + \frac2x \mathcal{A}'(x) + \left[\kappa^2 - \frac2{x^2}\right] \mathcal{A}(x) = x \tilde{\rho}(x), \label{eq:ODE-alpha}
\eeq  
where we have defined
\beq
\kappa \equiv \kcs R
\eeq 
and $\tilde{\rho}(x) \equiv \rho(Rx)/\overline{\rho}$ is the dimensionless density profile.
Outside the star, the right-hand-side of Eq.~(\ref{eq:ODE-alpha}) vanishes and $\mathcal{A}$ is a homogeneous solution:
\beq
\mathcal{A}(x \geq 1) = a ~ j_1(\kappa x) + b~ y_1(\kappa x), \label{eq:homogeneous-sol}
\eeq
where $a$ and $b$ are integration constants, and $j_n$ and $y_n$ are the order $n$ spherical Bessel functions of the first and second kind, respectively. Expanding Eq.~(\ref{eq:ODE-alpha}) near the origin, one finds that $\mathcal{A}(0) = 0$ and $\mathcal{A}$ must be at least linear in $x$ near $x = 0$. Denoting $\mathcal{A}_{\rm P}$ a particular solution of Eq.~(\ref{eq:ODE-alpha}), we therefore have
\beq
\mathcal{A}(x \leq 1) = \mathcal{A}_{\rm P}(x) + c ~ j_1(\kappa x),
\eeq
where we have eliminated the homogeneous solution proportional to $y_1$ as it diverges at $x= 0$.

We therefore have three integration constants $a, b, c$ to determine. Requiring $\mathcal{A}$ and its first derivative to be continuous at the boundary of the star [as is required by the finiteness of all the coefficients in Eq.~(\ref{eq:ODE-alpha})] provides only two constraints. The homogeneous solution (\ref{eq:homogeneous-sol}) is well behaved at infinity and one cannot, a priori, eliminate any of the constants $a, b$ or combination thereof.

This issue hints at potentially deeper problems with the theory, which has not yet been shown to be a well posed initial value problem (see however Ref.~\cite{Grumiller_08} which derived the boundary terms necessary to render the Dirichlet problem well defined). It is also possible that in a fully time-dependent study, the evolution selects a unique homogeneous solution outside the source. Since such a study is well beyond the scope of this work, and any choice of homogeneous solution outside the source does not affect any of our constraints quantitatively, we follow SE08 and choose the \emph{ansatz} $a = 0$ in Eq.~(\ref{eq:homogeneous-sol}). SE08 justify this choice by pointing out that in the limit $\kcs R \ll 1$ and $r \gtrsim R$, the function $j_1(\kcs r) \sim \frac 13 \kcs r$ increases with distance outside of the source, which they interpret as unphysical.

With this ansatz, we can solve explicitly for the constants $b$ and $c$ given a particular solution $\mathcal{A}_{\rm P}$ by requiring the continuity and smoothness of $\mathcal{A}$ at the boundary of the star. Solving the resulting linear second order algebraic equation, and using properties of the spherical Bessel functions, we arrive at
\barr
b &=& \kappa^2 j_2(\kappa) \mathcal{A}_{\rm P}(1) + \kappa j_1(\kappa) \left[\mathcal{A}_{\rm P}'(1) - \mathcal{A}_{\rm P}(1)\right], \label{eq:b-sol}\\
c &=& \kappa^2 y_2(\kappa) \mathcal{A}_{\rm P}(1) + \kappa y_1(\kappa) \left[\mathcal{A}_{\rm P}'(1) - \mathcal{A}_{\rm P}(1)\right]. \label{eq:c-sol}
\earr
Before proceeding with specific examples, we write down the CS correction to the gravitomagnetic field in the same form as in SE08:
\barr
\vec{B}_{\rm CS} &=& 4 \pi \overline{\rho} R^2 \Big{\{} D_1(r) \vec{\Omega} + D_2(r) \hat{r} \times \vec{\Omega} \nonumber\\
&& ~~~ ~~~~~~~~~ + D_3(r) \hat{r} \times (\hat{r} \times \vec{\Omega})\Big{\}}, 
\earr
where, using Eqs.~(\ref{eq:BCSphi})-(\ref{eq:BCStheta}) and (\ref{eq:separable}), the functions $D_i$ are given by (with $x = r/R$):
\barr
D_1(r) &=& 2 \frac{\mathcal{A}(x)}{x},\\
D_2(r) &=& \kappa \mathcal{A}(x), \\
D_3(r) &=& \frac{\mathcal{A}(x)}{x} - \mathcal{A}'(x). \label{eq:D3}
\earr
In particular, outside the star, we obtain, after simplifying Eq.~(\ref{eq:D3}):
\barr
D_1(r \geq R) &=& b~\frac{2 R}r  y_1(\kcs r), \label{eq:D1-general}\\
D_2(r \geq R) &=& b~ \kcs R ~y_1(\kcs r), \\
D_3(r \geq R) &=& b~\kcs R~y_2(\kcs r).\label{eq:D3-general}
\earr

For a constant density object, we have $\tilde{\rho}(x) = 1$ and the function $\mathcal{A}_{\rm P}(x) = x/\kappa^2$ is a particular solution of Eq.~(\ref{eq:ODE-alpha}). We therefore obtain $b = j_2(\kappa)$, $c = y_2(\kappa)$ form Eqs.~(\ref{eq:b-sol})-(\ref{eq:c-sol}). One can then easily check that our solution for the gravitomagnetic field is then in agreement with that found by SE08.

Assuming a constant density is a good approximation for the Earth, in which the density varies by only a factor of a few from the center to the edge. In a neutron star, however, the surface density is in general of the order of $\sim 10^{-7}$ times the mean density, and the density profile varies smoothly from the core to the edge. This is important as a smoother edge will result in a smaller amplitude for the high spatial frequency modes $\kcs R \gg 1$, as we shall see below.

As a simple yet more realistic model for a neutron star, we assume a density profile $\rho(r) = \rho_c \left[ 1 - (r/R)^2\right]$. The dimensionless density profile is then
\beq
\tilde{\rho}(x) = \frac52 (1 - x^2),
\eeq
where the normalization ensures that the average density is $\overline{\rho}$. A particular solution for Eq.~(\ref{eq:ODE-alpha}) with this density profile is
\beq
\mathcal{A}_{\rm P}(x) = 25 \frac{x}{\kappa^4} + \frac52 \frac{x(1- x^2)}{\kappa^2}.
\eeq
Using this solution into Eqs.~(\ref{eq:b-sol})-(\ref{eq:c-sol}) and after simplifications, we obtain 
\beq
b =5 \frac{j_3(\kappa)}{\kappa}, \ \ \  \ c = 5 \frac{y_3(\kappa)}{\kappa}. \label{eq:b-realistic}
\eeq
Using Eq.~(\ref{eq:b-realistic}) into Eqs.~(\ref{eq:D1-general})-(\ref{eq:D3-general}), we obtain Eqs.~(\ref{eq:D1-main})-(\ref{eq:D3-main}) for the CS gravitomagnetic field outside a rotating neutron star.

\section{Derivation of the rate of change of orbital elements}\label{app:orbital}

This appendix is dedicated to explicitly derive the rate of periastron precession given in Eq.~(\ref{eq:omega_dot}) in the case where the orbital angular momentum is aligned with the spin vector of the central rotating object. We also show that the poloidal gravitomagnetic field leads to very interesting dynamics. We start with the general Gaussian perturbation equations. 

\subsection{Gaussian perturbation equations}

The gaussian perturbation equations (see for example Ref.~\cite{Roy}) give the rate of change of osculating constants of motion $c_i$ for a test mass evolving in a newtonian potential and subject to a perturbing force per unit mass $\delta \vec{f}$:
\beq
\frac{d c_i}{d t} = \frac{\partial c_i}{\partial \vec{v}}\cdot \delta \vec{f}. \label{eq:gauss}
\eeq
Denoting $\mu$ the total mass of the system, a trajectory in a newtonian potential is defined by 6 constants of motion: the energy per unit mass $E = \frac12 v^2 - \mu/r$, the angular momentum per unit mass $\vec{\ell} \equiv \vec{r} \times \vec{v}$ and the eccentricity vector $\vec{e} \equiv \frac{1}{\mu}\vec{v}\times \vec{\ell} - \hat{r}$. Note that $\vec{\ell}$ and $\vec{e}$ together contain only 5 constants of motion as $\vec{\ell}\cdot \vec{e} = 0$. The eccentricity vector has magnitude $e$, the eccentricity of the orbit, and points along the symmetry axis of the trajectory, towards the pericenter. Using Eq.~(\ref{eq:gauss}), we obtain that the rates of change of the constants of motion are
\barr
\frac{d E}{d t} &=& \vec{v} \cdot \delta \vec{f}, \label{eq:energy}\\
\frac{d \vec{\ell}}{d t} &=& \vec{r} \times \delta \vec{f},\\
\frac{d \vec{e}}{dt} &=& \frac1{\mu} \left[2 (\vec{v}\cdot \delta \vec{f}) \vec{r} -  (\vec{v}\cdot \vec{r}) \delta \vec{f} - (\vec{r} \cdot \delta \vec{f}) \vec{v} \right].
\earr

\subsection{Application to a general gravitomagnetic force}
The gravitomagnetic force is $\delta \vec{f} = - 4 \vec{v} \times \vec{B}$. This force does not do any work and therefore the energy (and as a consequence the semi-major axis $a$) is conserved, as can be seen from Eq.~(\ref{eq:energy}). The gaussian perturbation equations for the angular momentum and eccentricity become:
\barr
\frac{d \vec{\ell}}{dt} &=& 4(\vec{r}\cdot \vec{v}) \vec{B} - 4 (\vec{r}\cdot \vec{B}) \vec{v}, \label{eq:ldot-B}\\
\frac{d \vec{e}}{dt} &=&  \frac4{\mu} \left[(\vec{v}\cdot \vec{r}) (\vec{v}\times \vec{B})+ (\vec{B}\cdot \vec{\ell})\vec{v}\right]. \label{eq:edot-B}
\earr
We now consider a general gravitomagnetic field of the form
\beq
\vec{B} = B_1(r) \hat{\Omega} + B_2(r) \hat{r} \times \hat{\Omega} + B_3(r) \hat{r} \times (\hat{r} \times \hat{\Omega}), \label{eq:B-general}
\eeq
where $\hat{\Omega}$ is a fixed unit vector. Eq.~(\ref{eq:B-general}) describes both the GR gravitomagnetic field (with $B_1(r) = J/r^3, \ B_2(r) = 0,  \ B_3(r) = \frac32 J/r^3$, where $J$ is the spin of the rotating central object) and the CS gravitomagnetic field, Eq.~(\ref{eq:BCS}).

We start by considering the simple case where $\hat{\Omega}~ || ~\vec{\ell}$, which is the relevant case for the binary pulsar. We do so for a general field with no poloidal component ($B_2 = 0$). We then separately consider the effect of the poloidal component for a general relative orientation of $\hat{\Omega}$ and $\hat{\ell}$.

\subsection{Case of $\hat{\Omega}~|| ~\vec{\ell}$, no poloidal component.}

In that case the gravitomagnetic field simplifies to $\vec{B} = [B_1(r) - B_3(r)] \hat{\Omega} \equiv B_0(r) \hat{\ell}$. Eqs.~(\ref{eq:ldot-B}) and (\ref{eq:edot-B}) become (using $\vec{r} \cdot \vec{v} = r \dot{r}$):
\barr
\frac{d \vec{\ell}}{dt} &=& 4r \dot{r} B_0(r) \hat{\ell}, \label{eq:ldot-case1}\\
\frac{d \vec{e}}{dt} &=&  \frac4{\mu} B_0(r)\left[r \dot{r}  (\vec{v}\times \hat{\ell})+ \ell \vec{v}\right]. \label{eq:edot-case1}
\earr
The right-hand-side of Eq.~(\ref{eq:ldot-case1}) is a total time derivative and there are therefore no secular changes of the angular momentum vector: 
\beq
\Big{\langle} \frac{d \vec{\ell}}{dt} \Big{\rangle}_{T} = \vec{0},
\eeq
where 
\beq
\langle X \rangle_{\rm T} \equiv \frac1T \int_{0}^T X(t) dt
\eeq
is the time-average of $X$ over one orbital period $T$. Since $\ell^2 = \mu a (1 - e^2)$, no secular changes of $\vec{\ell}$ (and therefore of its magnitude $\ell$) imply that there are no secular changes in the eccentricity $e \equiv ||\vec{e}||$. As a consequence, $\langle d \vec{e}/dt \rangle\cdot \vec{e} = 0$ [this can be also checked explicitly from Eq.~(\ref{eq:edot-case1})]. Since the right-hand side of Eq.~(\ref{eq:edot-case1}) lies in the orbital plane, we conclude that the secular precession of the eccentricity vector must be of the form
\beq
\Big{\langle} \frac{d \hat{e}}{dt} \Big{\rangle}_{T} = \dot{\omega} ~\hat{u}, 
\eeq
where $\hat{u} \equiv \hat{\ell} \times \hat{e}$. To get the rate of periastron precession $\dot{\omega}$, we need to project Eq.~(\ref{eq:edot-case1}) on the unit vector $\hat{u}$. Before doing so, it is useful to recall some properties of an elliptic orbit. The radial coordinate $r$ is given by 
\beq
r(f) \equiv \frac{p}{1 + e \cos f},\label{eq:r(f)}
\eeq
where $p \equiv a(1 - e^2)$ is the semi-latus rectum and $f$ is the true anomaly, the angle between $\vec{e}$ and $\vec{r}$. Using conservation of angular momentum, $\ell = r^2 \dot{f}$, the derivative of $r$ with respect to time can be written as 
\beq
\dot{r} = \frac{\ell}{p} e \sin f.
\eeq 
Finally, the projections of the velocity on the orthonormal vectors $\hat{e}, ~\hat{u}$ are given by
\beq
\vec{v} \cdot \hat{e} = -\frac{\ell}p \sin f, \ \ \ \ \vec{v} \cdot \hat{u} = \frac{\ell}{p}(e + \cos f).
\eeq
With these relations at hand and after simplifying, we finally obtain the rate of periastron precession:
\barr
&&\dot{\omega} = 4\Big{\langle} \frac r p B_0(r)  (2 + e^{-1}(1+e^2) \cos f) \Big{\rangle}_{T}\nonumber\\
&& = \frac4{(1 - e^2)^{3/2}} \Bigg{\langle} \frac{r^3}{a^3} B_0(r)  \left(2 + \frac{1 + e^2}e \cos f\right) \Bigg{\rangle}_{f},~~~ 
\earr
where in the second line we have converted the time average to an angular average
\beq
\langle X\rangle_f \equiv \frac1{2\pi} \int_0^{2\pi} X(f) d f. \label{eq:<X>f}
\eeq
We can apply this result to standard GR, with $B_0(r) = - \frac12 J/r^3$, and obtain the well known expression for the Lense-Thirring drag [note that this is the total rate of change of the longitude of the pericenter which is the sum of the rates of precession of the longitude of the ascending node and the argument of pericenter]:
\beq
\dot{\omega}_{\rm GR} = -\frac{4 J}{a^3(1 - e^2)^{3/2}}. 
\eeq
For a slowly varying gravitomagnetic field, we can find the leading order of the precession rate in eccentricity, for $e \ll 1$, by expanding $r \approx a - a e \cos f$. We find
\beq
\dot{\omega} = 2 [B_0(a) - a B_0'(a)] + \mathcal{O}(e).
\eeq

\subsection{Case of a poloidal gravitomagnetic field}\label{app:poloidal}

We now consider the case of a purely poloidal gravitomagnetic field, $\vec{B} = B(r) \hat{r} \times \hat{\Omega}$, for arbitrary orientation of the vector $\hat{\Omega}$. This case is very interesting as it is completely absent in standard GR and, as we shall see, leads to unique effects. Using Eq.~(\ref{eq:ldot-B}), we find
\barr
&&\frac{d \vec{\ell}}{dt} = 4 r \dot{r} B(r) \hat{r} \times \hat{\Omega}\nonumber\\
&&= 4 r \frac{\ell}{p} e \sin f B(r) \left[ \cos f (\hat{e} \times \hat{\Omega}) + \sin f (\hat{u} \times \hat{\Omega})\right].~~~
\earr
The first term inside the brackets averages to zero (it is odd under the change $f \rightarrow -f$), and we therefore obtain
\beq
\frac1{\ell}\Big{\langle} \frac{d \vec{\ell}}{dt} \Big{\rangle}_{T} = \gamma_1 e \hat{u} \times \hat{\Omega},
\eeq
where we have defined the rate
\beq
\gamma_1 \equiv \frac{4}{p} \langle r B(r) \sin^2 f \rangle_T.  
\eeq
Note that $\gamma_1 = 2B(a)$ to lowest order in eccentricity.
Projecting this equation on $\hat{\ell}$, we obtain that the \emph{magnitude} of the orbital angular momentum changes with a secular rate
\beq
\Big{\langle} \frac{d \ell}{d t}  \Big{\rangle}_{T}= - \gamma_1 e (\hat{\Omega} \cdot \hat{e}) \ell.
\eeq
This result is in stark contrast with GR, where (at least to lowest order) spin-orbit coupling does not change the magnitude of the orbital angular momentum or the spin, but only their relative orientation. We moreover obtain that the \emph{orientation} of $\vec{\ell}$ changes with the rate
\beq
\Big{\langle} \frac{d \hat{\ell}}{d t}  \Big{\rangle}_{T} = \gamma_1 e (\hat{\Omega} \cdot \hat{\ell}) \hat{e}.
\eeq
We can now readily obtain the rate of change of the eccentricity, using $\ell^2 = \mu a ( 1- e^2)$:
\beq
\Big{\langle} \frac{d e}{d t}  \Big{\rangle}_{T} = \gamma_1 (1 - e^2)(\hat{\Omega} \cdot \hat{e}).
\eeq\\
We see that circular orbits are unstable: a tiny eccentricity grows with rate $2 B(a) (\hat{\Omega} \cdot \hat{e})$. To obtain the rate of change of the orientation of the eccentricity vector, we first notice that $\hat{e}\cdot \hat{\ell} = 0$ at all times, which gives us
\beq
\Big{\langle} \frac{d \hat{e}}{d t}  \Big{\rangle}_{T} \cdot \hat{\ell} = -\Big{\langle} \frac{d \hat{\ell}}{d t}  \Big{\rangle}_{T} \cdot \hat{e} = - \gamma_1  e (\hat{\Omega} \cdot \hat{\ell}). 
\eeq
Our last and most tedious task is to project Eq.~(\ref{eq:edot-B}) on $\hat{u} = \hat{\ell} \times \hat{e}$. After some manipulations, we obtain
\beq
\Big{\langle} \frac{d \hat{e}}{d t}  \Big{\rangle}_{T} \cdot \hat{u} = \frac{\gamma_2}e (\hat{\Omega} \cdot \hat{u}), 
\eeq
where we have defined the rate
\beq
\gamma_2 \equiv \frac4p \langle r B(r) (2e \cos f + (1 + e^2) \cos^2 f)\rangle_T,
\eeq
which, to lowest order in eccentricity, has value $\gamma_2 = 2 B(a)$. Again, this indicates an instability: the eccentricity vector $\vec{e} = e \hat{e}$ rotates with a divergent angular rate $2B(a) e^{-1} (\hat{\Omega} \cdot \hat{u})$ for small eccentricities.

We have therefore shown that the poloidal component of the gravitomagnetic field leads to unique dynamics:

$(i)$ A rotation of $\vec{e}$ and $\vec{\ell}$ around $\hat{\ell} \times \hat{e}$, with angular rate $\gamma_1 e  (\hat{\Omega}\cdot \hat{\ell})$, vanishing for circular orbits.

$(ii)$ A precession of the eccentricity vector around $\vec{\ell}$ with angular rate $\gamma_2 e^{-1} (\hat{\Omega}\cdot \hat{u})$, divergent for arbitrarily small eccentricities.

$(iii)$ A change of the magnitude of the orbital angular momentum with rate $\dot{\ell}/\ell = - \gamma_1 e(\hat{\Omega}\cdot \hat{e})$, accompanied with a change of the eccentricity with rate $\dot{e} = \gamma_1 (1 - e^2) (\hat{\Omega}\cdot \hat{e})$, non-vanishing even for arbitrarily small eccentricities.

A full analysis of the dynamics of the system would require accounting for the standard GR precession effects, spin-orbit coupling, etc..., as well as a detailed analysis of the gravitomagnetic field induced by the orbital motion itself. This will be the subject of future work.

\section{Confidence level on the constraint}\label{app:constraints}
Equation (\ref{eq:omega.dot.final}) is of the form 
\barr
\dot{\omega}_{\rm CS} = \alpha \kcs^{-5.5} \sin(\kcs a)\cos(\kcs a e + \frac{\pi}4)\nonumber\\
~~~~~~~ \times \cos(\kcs R_A) \sin (\kcs R_B)\label{eq:omega.dot.1},
\earr
where $\alpha$ is a numerical constant depending on the system considered. Assuming $R_A = R_B \equiv R$, we simplify this expression to
\beq
\dot{\omega}_{\rm CS} = \frac{\alpha}2 \kcs^{-5.5} \mathcal{S}_3,
\eeq 
where $\mathcal{S}_3$ is the product of the 3 sinusoidal functions
\beq
\mathcal{S}_3 \equiv \sin(\kcs a)\cos(\kcs a e + \frac{\pi}4) \sin(2 \kcs R).
\eeq
From now on, to simplify the notation, we will use the CS characteristic length $L \equiv \kcs^{-1}$. $L=0$ corresponds to standard GR, and the CS-induced periastron precession rate is $\dot{\omega}_{\rm CS} = \frac{\alpha}2 L^{5.5} \mathcal{S}_3$. It will also be convenient to define the lengthscale $L_0 \equiv (2 \delta \dot{\omega}/\alpha)^{1/5.5}$, so that one can rewrite 
\beq
\dot{\omega}_{\rm CS} = \mathcal{S}_3 \left(\frac{L}{L_0}\right)^{5.5} \delta \dot \omega.
\eeq
In Sec.~\ref{sec:new constraints} we have, as a first approximation, simply used the envelope of Eq.~(\ref{eq:omega.dot.1}) $|\dot{\omega}_{\rm CS}| \sim \alpha L^{5.5}$, to derive constraints on $L$ from observational constraints $|\dot{\omega}_{\rm CS}| \lesssim \delta \dot{\omega}$. Clearly, however, there are always arbitrarily large values of $L$ that will satisfy any observational constraint, for example $L = N \pi a^{-1}$ for any integer $N$, or values close enough from this. One should precisely quantify this caveat and rather quote constraints in the form ``$L < L_{\rm max}$, with $X$\% confidence'', meaning that there is still a probability $1-X\%$ that $|\dot{\omega}_{\rm CS}| < \delta \dot{\omega}$ for $L \approx L_{\rm max}$.

\begin{figure}
\includegraphics[width = 80 mm]{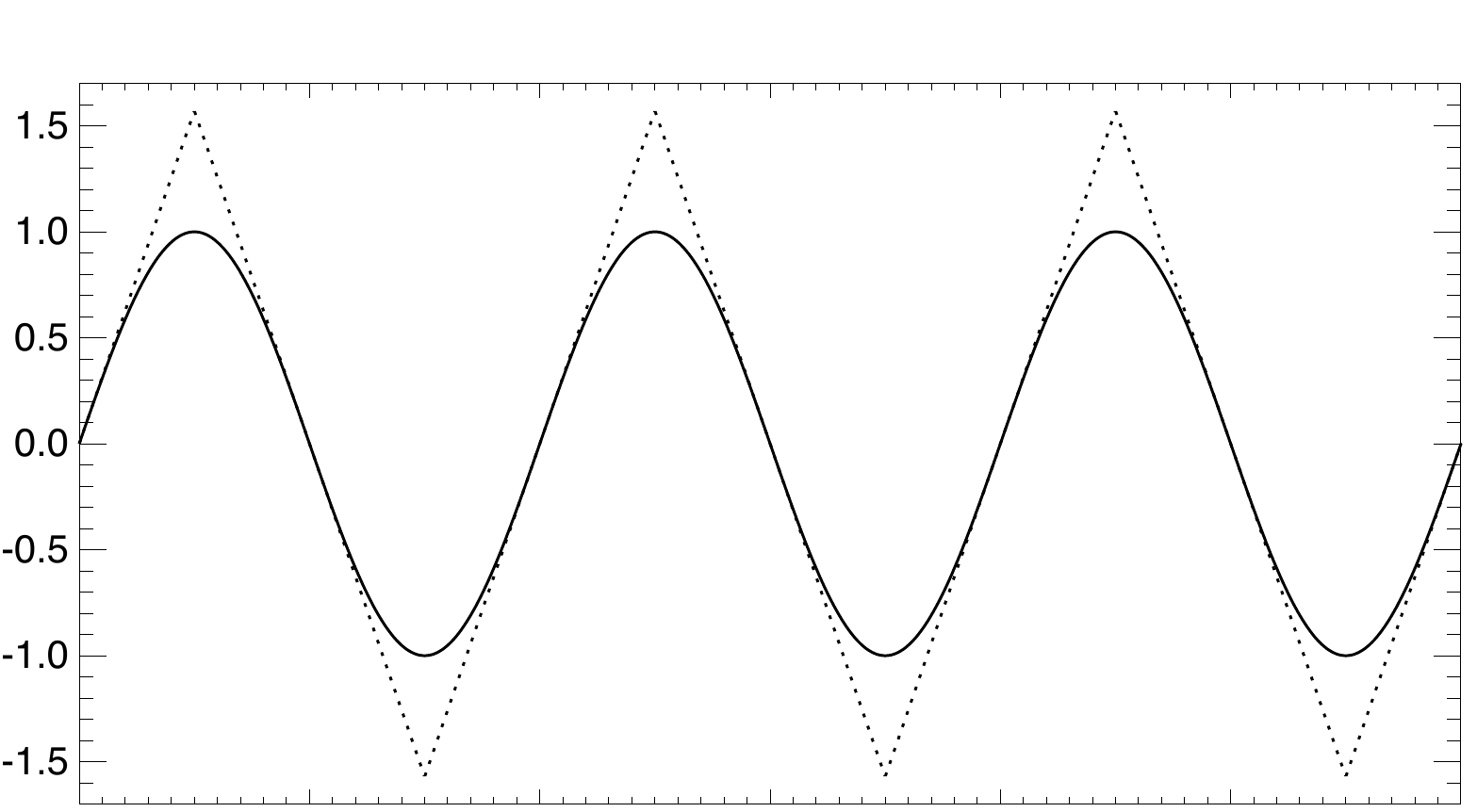}
\caption{To simplify the calculation, we approximate each sinusoidal function with a triangular periodic function with the same tangents near zeros.} \label{fig:triangles} 
\end{figure}

In order to do so, we need to evaluate the probability distribution for the magnitude of the product of the three sinusoidal functions $\mathcal{S}_3$. We first notice that the three functions can be assumed to be uncorrelated, because of the largely different scales of their arguments ($R \ll ae \ll a$). Our first task is therefore to find the probability distribution for the amplitude of a single sinusoidal function. Since we are mainly concerned about intervals where this amplitude might be small, we approximate each sinusoidal function with a triangular periodic function with the same tangents near zeros (see Fig.~\ref{fig:triangles}). With this approximation, the probability that any single sinusoidal function has an absolute value in the range $[\epsilon, \epsilon + d \epsilon]$ is $p_1(\epsilon) d \epsilon = \frac{2}{\pi} d \epsilon$ (this is correct to order $\mathcal{O}(\epsilon^2)$ for $\epsilon \ll 1$ but we formally use it for $\epsilon \in [0, \pi/2]$). We now would like to compute the probability distribution $p_3(\epsilon \equiv |\mathcal{S}_3|)$ for the absolute value of the product of the $3$ uncorrelated sinusoidal functions. It is easier to work with $u_i \equiv \ln (2 \epsilon_i/\pi)$, where $\epsilon_i \equiv|\sin(...)|$ is the absolute value of one of the three sinusoidal functions. We have $u = \ln (2 \epsilon/\pi) = u_1 + u_2 + u_3$, where a change of variables gives us $p_1(u_i) = 2 \epsilon_i /\pi= \exp(u_i)$. Therefore, we get
\barr
p_3(u) &=& \int_{u}^{0} p_1(u_1) d u_1 \int_{u-u_1}^{0} p_1(u_2) d u_2 p_1(u - u_1 - u_2)\nonumber\\
&=& \frac12 u^2 \exp(u).\label{eq:p3(u)}
\earr
We can now obtain the probability that the product of three sinusoidal functions is less than a certain value $\epsilon$:
\barr
&&P(|\mathcal{S}_3|<\epsilon) = \int_0^{\epsilon} p_3(\epsilon') d \epsilon' \nonumber\\
 &&= \frac{2 \epsilon}{\pi}\left( 1 + |\ln(2 \epsilon/\pi)| + \frac12 [\ln(2 \epsilon/\pi)]^2\right), \label{eq:PS3}
\earr 
where we have used Eq.~(\ref{eq:p3(u)}) for $p_3(u')$ and replaced $u = \ln(2 \epsilon/\pi)$ in the final result. Because of the $(\ln \epsilon)^2$ term, we see that  there is actually quite a significant probability that the product has a small amplitude.
For example, there is a 48\% chance that the product of sinusoidal functions is less than 0.1 in magnitude and still a 12\% chance that it is less than 0.01.

\begin{figure}
\includegraphics[width = 85 mm]{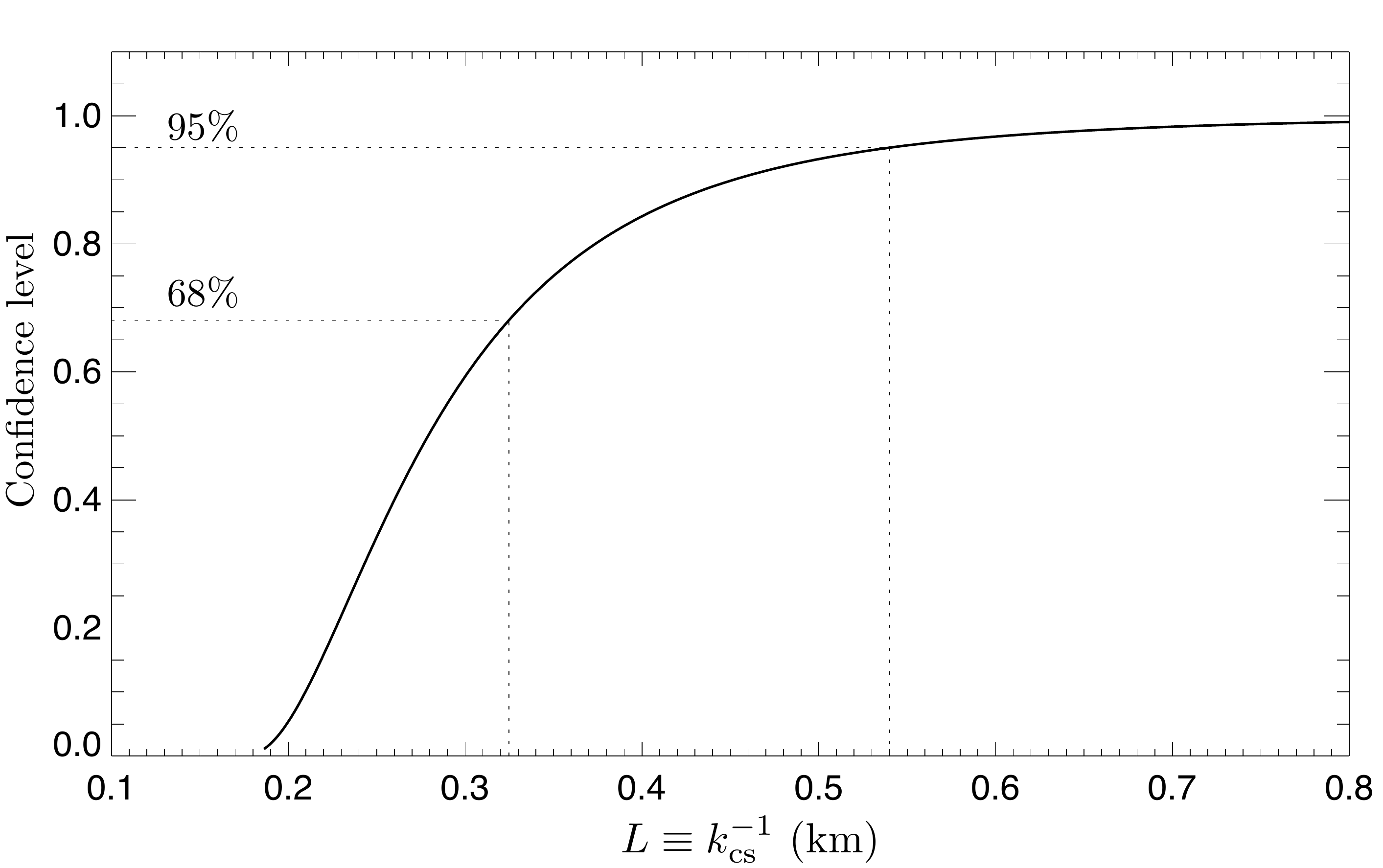}
\caption{Probability that the observational constraint $|\dot{\omega}_{\rm CS}| < \delta \dot{\omega}$ is violated as a function of the CS lengthscale $\kcs^{-1}$.} \label{fig:confidence} 
\end{figure}

We can now precisely quantify how reliable a constraint is. For a given $L$, $|\dot{\omega}_{\rm CS}| < \delta \dot{\omega}$ is equivalent to $|\mathcal{S}_3| < (L_0/L)^{5.5}$, the probability of which is given by Eq.~(\ref{eq:PS3}) with $\epsilon =(L_0/L)^{5.5}$. We show the probability $1 - P(|\mathcal{S}_3| < (L_0/L)^{5.5})$ as a function of $L$ in Fig.~\ref{fig:confidence}. We interpret this probability as a level of confidence on the constraint $\kcs^{-1} < L$. For example, we obtain
\barr
\kcs^{-1} &<& 0.32 \ \textrm{km} \ \ \ [68 \% \textrm{ confidence}], \nonumber\\
\kcs^{-1} &<& 0.54 \ \textrm{km} \ \ \ [95 \%\textrm{ confidence}]. \nonumber
\earr
Even though here we have chosen the often used 68\% and 95\% confidence intervals, we emphasize that the probability distribution is not a gaussian. Moreover, we insist that the strict meaning of these confidence intervals is that one may still have $\kcs^{-1}$ equal to the maximum value quoted with the probability complementary to the confidence level.


\newpage

\bibliography{chern_refs}

%Merlin.mbs v4.21 2009-07-09.
\begin{thebibliography}{10}%
\makeatletter
\providecommand \@ifxundefined [1]{%
 \ifx #1\undefined \expandafter \@firstoftwo
 \else \expandafter \@secondoftwo
\fi
}%
\providecommand \@ifnum [1]{%
 \ifnum #1\expandafter \@firstoftwo
 \else \expandafter \@secondoftwo
\fi
}%
\providecommand \enquote [1]{``#1''}%
\providecommand \bibnamefont  [1]{#1}%
\providecommand \bibfnamefont [1]{#1}%
\providecommand \citenamefont [1]{#1}%
\providecommand\href[0]{\@sanitize\@href}%
\providecommand\@href[1]{\endgroup\@@startlink{#1}\endgroup\@@href}%
\providecommand\@@href[1]{#1\@@endlink}%
\providecommand \@sanitize [0]{\begingroup\catcode`\&12\catcode`\#12\relax}%
\@ifxundefined \pdfoutput {\@firstoftwo}{%
 \@ifnum{\z@=\pdfoutput}{\@firstoftwo}{\@secondoftwo}%
}{%
 \providecommand\@@startlink[1]{\leavevmode\special{html:<a href="#1">}}%
 \providecommand\@@endlink[0]{\special{html:</a>}}%
}{%
 \providecommand\@@startlink[1]{%
  \leavevmode
  \pdfstartlink
   attr{/Border[0 0 1 ]/H/I/C[0 1 1]}%
   user{/Subtype/Link/A<</Type/Action/S/URI/URI(#1)>>}%
  \relax
 }%
 \providecommand\@@endlink[0]{\pdfendlink}%
}%
\providecommand \url  [0]{\begingroup\@sanitize \@url }%
\providecommand \@url [1]{\endgroup\@href {#1}{\urlprefix}}%
\providecommand \urlprefix [0]{URL }%
\providecommand \Eprint[0]{\href }%
\@ifxundefined \urlstyle {%
  \providecommand \doi [1]{doi:\discretionary{}{}{}#1}%
}{%
  \providecommand \doi [0]{doi:\discretionary{}{}{}\begingroup
  \urlstyle{rm}\Url }%
}%
\providecommand \doibase [0]{http://dx.doi.org/}%
\providecommand \Doi[1]{\href{\doibase#1}}%
\providecommand \bibAnnote [3]{%
  \BibitemShut{#1}%
  \begin{quotation}\noindent
    \textsc{Key:}\ #2\\\textsc{Annotation:}\ #3%
  \end{quotation}%
}%
\providecommand \bibAnnoteFile [2]{%
  \IfFileExists{#2}{\bibAnnote {#1} {#2} {\input{#2}}}{}%
}%
\providecommand \typeout [0]{\immediate \write \m@ne }%
\providecommand \selectlanguage [0]{\@gobble}%
\providecommand \bibinfo [0]{\@secondoftwo}%
\providecommand \bibfield [0]{\@secondoftwo}%
\providecommand \translation [1]{[#1]}%
\providecommand \BibitemOpen[0]{}%
\providecommand \bibitemStop [0]{}%
\providecommand \bibitemNoStop [0]{.\EOS\space}%
\providecommand \EOS [0]{\spacefactor3000\relax}%
\providecommand \BibitemShut [1]{\csname bibitem#1\endcsname}%
%</preamble>
\bibitem{Will_06}%
  \BibitemOpen
  \bibfield{author}{%
  \bibinfo {author} {\bibfnamefont{C.~M.}\ \bibnamefont{Will}},\ }%
  \bibfield{journal}{%
  \bibinfo {journal} {Living Rev. Relativity}\ }%
  \textbf{\bibinfo {volume} {9}},\ \bibinfo {pages} {3} (\bibinfo {year}
  {2006})%
  \bibAnnoteFile{NoStop}{Will_06}%
\bibitem{Jackiw_Pi_03}%
  \BibitemOpen
  \bibfield{author}{%
  \bibinfo {author} {\bibfnamefont{R.}~\bibnamefont{{Jackiw}}}\ and\ \bibinfo
  {author} {\bibfnamefont{S.}~\bibnamefont{{Pi}}},\ }%
  \bibfield{journal}{%
  \Doi{10.1103/PhysRevD.68.104012}{\bibinfo {journal} {\prd}}\ }%
  \textbf{\bibinfo {volume} {68}},\ \bibinfo {pages} {104012} (\bibinfo {year}
  {2003})%
  \bibAnnoteFile{NoStop}{Jackiw_Pi_03}%
\bibitem{Alexander_Yunes_09}%
  \BibitemOpen
  \bibfield{author}{%
  \bibinfo {author} {\bibfnamefont{S.}~\bibnamefont{{Alexander}}}\ and\
  \bibinfo {author} {\bibfnamefont{N.}~\bibnamefont{{Yunes}}},\ }%
  \bibfield{journal}{%
  \Doi{10.1016/j.physrep.2009.07.002}{\bibinfo {journal} {\physrep}}\ }%
  \textbf{\bibinfo {volume} {480}},\ \bibinfo {pages} {1} (\bibinfo {year}
  {2009})%
  \bibAnnoteFile{NoStop}{Alexander_Yunes_09}%
\bibitem{Lue_Wang_99}%
  \BibitemOpen
  \bibfield{author}{%
  \bibinfo {author} {\bibfnamefont{A.}~\bibnamefont{{Lue}}}, \bibinfo {author}
  {\bibfnamefont{L.}~\bibnamefont{{Wang}}},\ and\ \bibinfo {author}
  {\bibfnamefont{M.}~\bibnamefont{{Kamionkowski}}},\ }%
  \bibfield{journal}{%
  \bibinfo {journal} {\prl}\ }%
  \textbf{\bibinfo {volume} {83}},\ \bibinfo {pages} {1506} (\bibinfo {year}
  {1999})%
  \bibAnnoteFile{NoStop}{Lue_Wang_99}%
\bibitem{Alexander_Peskin_06}%
  \BibitemOpen
  \bibfield{author}{%
  \bibinfo {author} {\bibfnamefont{S.~H.}\ \bibnamefont{{Alexander}}}, \bibinfo
  {author} {\bibfnamefont{M.~E.}\ \bibnamefont{{Peskin}}},\ and\ \bibinfo
  {author} {\bibfnamefont{M.~M.}\ \bibnamefont{{Sheikh-Jabbari}}},\ }%
  \bibfield{journal}{%
  \bibinfo {journal} {\prl}\ }%
  \textbf{\bibinfo {volume} {96}},\ \bibinfo {pages} {081301} (\bibinfo {year}
  {2006})%
  \bibAnnoteFile{NoStop}{Alexander_Peskin_06}%
\bibitem{Alexander_Yunes_07D}%
  \BibitemOpen
  \bibfield{author}{%
  \bibinfo {author} {\bibfnamefont{S.}~\bibnamefont{{Alexander}}}\ and\
  \bibinfo {author} {\bibfnamefont{N.}~\bibnamefont{{Yunes}}},\ }%
  \bibfield{journal}{%
  \Doi{10.1103/PhysRevD.75.124022}{\bibinfo {journal} {\prd}}\ }%
  \textbf{\bibinfo {volume} {75}},\ \bibinfo {pages} {124022} (\bibinfo {year}
  {2007})%
  \bibAnnoteFile{NoStop}{Alexander_Yunes_07D}%
\bibitem{Alexander_Yunes_07L}%
  \BibitemOpen
  \bibfield{author}{%
  \bibinfo {author} {\bibfnamefont{S.}~\bibnamefont{{Alexander}}}\ and\
  \bibinfo {author} {\bibfnamefont{N.}~\bibnamefont{{Yunes}}},\ }%
  \bibfield{journal}{%
  \Doi{10.1103/PhysRevLett.99.241101}{\bibinfo {journal} {\prl}}\ }%
  \textbf{\bibinfo {volume} {99}},\ \bibinfo {pages} {241101} (\bibinfo {year}
  {2007})%
  \bibAnnoteFile{NoStop}{Alexander_Yunes_07L}%
\bibitem{Smith_08}%
  \BibitemOpen
  \bibfield{author}{%
  \bibinfo {author} {\bibnamefont{[SE08]}}, \bibinfo {author}
  {\bibfnamefont{T.~L.}\ \bibnamefont{{Smith}}}, \bibinfo {author}
  {\bibfnamefont{A.~L.}\ \bibnamefont{{Erickcek}}}, \bibinfo {author}
  {\bibfnamefont{R.~R.}\ \bibnamefont{{Caldwell}}},\ and\ \bibinfo {author}
  {\bibfnamefont{M.}~\bibnamefont{{Kamionkowski}}},\ }%
  \bibfield{journal}{%
  \Doi{10.1103/PhysRevD.77.024015}{\bibinfo {journal} {\prd}}\ }%
  \textbf{\bibinfo {volume} {77}},\ \bibinfo {pages} {024015} (\bibinfo {year}
  {2008})%
  \bibAnnoteFile{NoStop}{Smith_08}%
\bibitem{Lageos}%
  \BibitemOpen
  \bibfield{author}{%
  \bibinfo {author} {\bibfnamefont{I.}~\bibnamefont{{Ciufolini}}}\ and\
  \bibinfo {author} {\bibfnamefont{E.~C.}\ \bibnamefont{{Pavlis}}},\ }%
  \bibfield{journal}{%
  \Doi{10.1038/nature03007}{\bibinfo {journal} {\nat}}\ }%
  \textbf{\bibinfo {volume} {431}},\ \bibinfo {pages} {958} (\bibinfo {year}
  {2004})%
  \bibAnnoteFile{NoStop}{Lageos}%
\bibitem{Yunes_Spergel_09}%
  \BibitemOpen
  \bibfield{author}{%
  \bibinfo {author} {\bibnamefont{[YS09]}}, \bibinfo {author}
  {\bibfnamefont{N.}~\bibnamefont{{Yunes}}},\ and\ \bibinfo {author}
  {\bibfnamefont{D.~N.}\ \bibnamefont{{Spergel}}},\ }%
  \bibfield{journal}{%
  \Doi{10.1103/PhysRevD.80.042004}{\bibinfo {journal} {\prd}}\ }%
  \textbf{\bibinfo {volume} {80}},\ \bibinfo {pages} {042004} (\bibinfo {year}
  {2009})%
  \bibAnnoteFile{NoStop}{Yunes_Spergel_09}%
\bibitem{Burgay_03}%
  \BibitemOpen
  \bibfield{author}{%
  \bibinfo {author} {\bibfnamefont{M.}~\bibnamefont{{Burgay}}} \emph{et~al.},\
  }%
  \bibfield{journal}{%
  \Doi{10.1038/nature02124}{\bibinfo {journal} {\nat}}\ }%
  \textbf{\bibinfo {volume} {426}},\ \bibinfo {pages} {531} (\bibinfo {year}
  {2003})%
  \bibAnnoteFile{NoStop}{Burgay_03}%
\bibitem{Kramer_Wex_09}%
  \BibitemOpen
  \bibfield{author}{%
  \bibinfo {author} {\bibfnamefont{M.}~\bibnamefont{{Kramer}}}\ and\ \bibinfo
  {author} {\bibfnamefont{N.}~\bibnamefont{{Wex}}},\ }%
  \bibfield{journal}{%
  \Doi{10.1088/0264-9381/26/7/073001}{\bibinfo {journal} {Class. Quantum
  Grav.}}\ }%
  \textbf{\bibinfo {volume} {26}},\ \bibinfo {pages} {073001} (\bibinfo {year}
  {2009})%
  \bibAnnoteFile{NoStop}{Kramer_Wex_09}%
\bibitem{Kramer_06}%
  \BibitemOpen
  \bibfield{author}{%
  \bibinfo {author} {\bibfnamefont{M.}~\bibnamefont{{Kramer}}} \emph{et~al.},\
  }%
  \bibfield{journal}{%
  \Doi{10.1126/science.1132305}{\bibinfo {journal} {Science}}\ }%
  \textbf{\bibinfo {volume} {314}},\ \bibinfo {pages} {97} (\bibinfo {year}
  {2006})%
  \bibAnnoteFile{NoStop}{Kramer_06}%
\bibitem{Effacing}%
  \BibitemOpen
  \bibfield{author}{%
  \bibinfo {author} {\bibfnamefont{T.}~\bibnamefont{{Damour}}},\ }%
  \bibfield{journal}{%
  \bibinfo {journal} {in {\em 300 Years of Gravitation}, Cambridge University
  Press, Edited by S.~Hawking and W.~Israel~}}%
   (\bibinfo {year} {1987})%
  \bibAnnoteFile{NoStop}{Effacing}%
\bibitem{Yunes_Pretorius_09}%
  \BibitemOpen
  \bibfield{author}{%
  \bibinfo {author} {\bibfnamefont{N.}~\bibnamefont{{Yunes}}}\ and\ \bibinfo
  {author} {\bibfnamefont{F.}~\bibnamefont{{Pretorius}}},\ }%
  \bibfield{journal}{%
  \Doi{10.1103/PhysRevD.79.084043}{\bibinfo {journal} {\prd}}\ }%
  \textbf{\bibinfo {volume} {79}},\ \bibinfo {pages} {084043} (\bibinfo {year}
  {2009})%
  \bibAnnoteFile{NoStop}{Yunes_Pretorius_09}%
\bibitem{Konno_09}%
  \BibitemOpen
  \bibfield{author}{%
  \bibinfo {author} {\bibfnamefont{K.}~\bibnamefont{{Konno}}}, \bibinfo
  {author} {\bibfnamefont{T.}~\bibnamefont{{Matsuyama}}},\ and\ \bibinfo
  {author} {\bibfnamefont{S.}~\bibnamefont{{Tanda}}},\ }%
  \bibfield{journal}{%
  \bibinfo {journal} {Progress of Theoretical Physics}\ }%
  \textbf{\bibinfo {volume} {122}},\ \bibinfo {pages} {561} (\bibinfo {year}
  {2009})%
  \bibAnnoteFile{NoStop}{Konno_09}%
\bibitem{Ferdman_08}%
  \BibitemOpen
  \bibfield{author}{%
  \bibinfo {author} {\bibfnamefont{R.~D.}\ \bibnamefont{{Ferdman}}}
  \emph{et~al.},\ }%
  \bibfield{journal}{%
  \Doi{10.1063/1.2900277}{\bibinfo {journal} {AIP Conference Series}}\ }%
  \textbf{\bibinfo {volume} {983}},\ \bibinfo {pages} {474} (\bibinfo {year}
  {2008})%
  \bibAnnoteFile{NoStop}{Ferdman_08}%
\bibitem{Will_1981}%
  \BibitemOpen
  \bibfield{author}{%
  \bibinfo {author} {\bibfnamefont{C.}~\bibnamefont{{Will}}},\ }%
  \bibfield{journal}{%
  \bibinfo {journal} {{\em Theory and experiment in gravitational physics},
  Cambridge University Press}}%
   (\bibinfo {year} {1981})%
  \bibAnnoteFile{NoStop}{Will_1981}%
\bibitem{Evan_thanks}%
  \BibitemOpen
  \bibinfo {journal} {We thank Evan O'Connor for providing neutron star density
  profiles for several masses and EOSs.}%
  \bibAnnoteFile{Stop}{Evan_thanks}%
\bibitem{LS_EOS}%
  \BibitemOpen
\bibfield{journal}{%
    }%
  \bibfield{author}{%
  \bibinfo {author} {\bibfnamefont{J.~M.}\ \bibnamefont{{Lattimer}}}\ and\
  \bibinfo {author} {\bibfnamefont{F.~D.}\ \bibnamefont{{Swesty}}},\ }%
  \bibfield{journal}{%
  \bibinfo {journal} {Nucl. Phys. A}\ }%
  \textbf{\bibinfo {volume} {535}},\ \bibinfo {pages} {331} (\bibinfo {year}
  {1991})%
  \bibAnnoteFile{NoStop}{LS_EOS}%
\bibitem{Shen_1}%
  \BibitemOpen
  \bibfield{author}{%
  \bibinfo {author} {\bibfnamefont{H.}~\bibnamefont{{Shen}}}, \bibinfo {author}
  {\bibfnamefont{H.}~\bibnamefont{{Toki}}}, \bibinfo {author}
  {\bibfnamefont{K.}~\bibnamefont{{Oyamatsu}}},\ and\ \bibinfo {author}
  {\bibfnamefont{K.}~\bibnamefont{{Sumiyoshi}}},\ }%
  \bibfield{journal}{%
  \Doi{10.1016/S0375-9474(98)00236-X}{\bibinfo {journal} {Nucl. Phys. A}}\ }%
  \textbf{\bibinfo {volume} {637}},\ \bibinfo {pages} {435} (\bibinfo {year}
  {1998})%
  \bibAnnoteFile{NoStop}{Shen_1}%
\bibitem{Shen_2}%
  \BibitemOpen
  \bibfield{author}{%
  \bibinfo {author} {\bibfnamefont{H.}~\bibnamefont{{Shen}}}, \bibinfo {author}
  {\bibfnamefont{H.}~\bibnamefont{{Toki}}}, \bibinfo {author}
  {\bibfnamefont{K.}~\bibnamefont{{Oyamatsu}}},\ and\ \bibinfo {author}
  {\bibfnamefont{K.}~\bibnamefont{{Sumiyoshi}}},\ }%
  \bibfield{journal}{%
  \Doi{10.1143/PTP.100.1013}{\bibinfo {journal} {Prog. Theor. Phys.}}\ }%
  \textbf{\bibinfo {volume} {100}},\ \bibinfo {pages} {1013} (\bibinfo {year}
  {1998})%
  \bibAnnoteFile{NoStop}{Shen_2}%
\bibitem{Grumiller_08}%
  \BibitemOpen
  \bibfield{author}{%
  \bibinfo {author} {\bibfnamefont{D.}~\bibnamefont{{Grumiller}}}, \bibinfo
  {author} {\bibfnamefont{R.}~\bibnamefont{{Mann}}},\ and\ \bibinfo {author}
  {\bibfnamefont{R.}~\bibnamefont{{McNees}}},\ }%
  \bibfield{journal}{%
  \Doi{10.1103/PhysRevD.78.081502}{\bibinfo {journal} {\prd}}\ }%
  \textbf{\bibinfo {volume} {78}},\ \bibinfo {pages} {081502} (\bibinfo {year}
  {2008})%
  \bibAnnoteFile{NoStop}{Grumiller_08}%
\bibitem{Roy}%
  \BibitemOpen
  \bibfield{author}{%
  \bibinfo {author} {\bibfnamefont{A.~E.}\ \bibnamefont{{Roy}}},\ }%
  \bibfield{journal}{%
  \bibinfo {journal} {Adam Hilger Ltd, Bristol}}%
   (\bibinfo {year} {1982})%
  \bibAnnoteFile{NoStop}{Roy}%
\end{thebibliography}%

\end{document}